\documentclass[12pt]{article}
\pdfoutput=1
\usepackage{putex}

\usepackage{amsmath, mathtools, amssymb}
\usepackage{graphicx,hyperref}
\usepackage{caption}
\usepackage{subcaption}
\usepackage{epstopdf}
\usepackage{enumerate}
\usepackage{cite}
\usepackage{tensor}
\usepackage{slashed}
\usepackage[aligntableaux=center]{ytableau}

\numberwithin{equation}{section}

\newcommand {\be} {\begin {equation}}
\newcommand {\ee} {\end {equation}}

\newcommand {\bes} {\begin {equation*}}
\newcommand {\ees} {\end {equation*}}

\newcommand{\Z}{\mathbb{Z}}

\newcommand{\R}{\mathbb{R}}
\newcommand{\C}{\mathbb{C}}

\def\Tr{\mop{Tr}}

\newcommand{\beq}{\begin{equation}}
\newcommand{\eeq}{\end{equation}}

\def\<{\langle}
\def\>{\rangle}

\newcommand{\bS}{\ensuremath{\mathbb{S}}}
\newcommand{\bT}{\ensuremath{\mathbb{T}}}

\newcommand{\cA}{\ensuremath{\mathcal{A}}}
\newcommand{\cB}{\ensuremath{\mathcal{B}}}
\newcommand{\cC}{\ensuremath{\mathcal{C}}}
\newcommand{\cD}{\ensuremath{\mathcal{D}}}

\newcommand{\cG}{\ensuremath{\mathcal{G}}}
\newcommand{\cH}{\ensuremath{\mathcal{H}}}
\newcommand{\cI}{\ensuremath{\mathcal{I}}}

\newcommand{\cL}{\ensuremath{\mathcal{L}}}
\newcommand{\cM}{\ensuremath{\mathcal{M}}}
\newcommand{\cN}{\ensuremath{\mathcal{N}}}
\newcommand{\cO}{\ensuremath{\mathcal{O}}}

\newcommand{\cQ}{\ensuremath{\mathcal{Q}}}
\newcommand{\cR}{\ensuremath{\mathcal{R}}}

\newcommand{\cT}{\ensuremath{\mathcal{T}}}

\newcommand{\cV}{\ensuremath{\mathcal{V}}}
\newcommand{\cW}{\ensuremath{\mathcal{W}}}
\newcommand{\cX}{\ensuremath{\mathcal{X}}}
\newcommand{\cY}{\ensuremath{\mathcal{Y}}}

\newcommand{\dd}{\mathrm{d}}
\newcommand{\rQ}{\mathrm{Q}}

\newcommand{\tq}{\widetilde{q}}
\newcommand{\tQ}{\widetilde{Q}}

\newcommand{\q}{\mathfrak{q}}

\newcommand\vertarrowbox[3][6ex]{%
	\begin{array}[t]{@{}c@{}} #2 \\
		\left\downarrow\vcenter{\hrule height #1}\right.\kern-\nulldelimiterspace\\
		\makebox[0pt]{#3}
	\end{array}%
}

\begin{document}

\institution{SCGP}{Simons Center for Geometry and Physics,\cr Stony Brook University, Stony Brook, NY 11794-3636, USA}

\title{On the 4d/3d/2d view of the SCFT/VOA correspondence}
\authors{Mykola Dedushenko\worksat{\SCGP}}

\abstract{We start with the SCFT/VOA correspondence formulated in the Omega-background approach, and connect it to the boundary VOA in 3d $\mathcal{N}=4$ theories and chiral algebras of 2d $\mathcal{N}=(0,2)$ theories. This is done using the dimensional reduction of the 4d theory on the topologically twisted and Omega-deformed cigar, performed in two steps. This paves the way for many more interesting questions, and we offer quite a few. We also use this approach to explain some older observations on the TQFTs produced from the generalized Argyres-Douglas (AD) theories reduced on the circle with a discrete twist. In particular, we argue that many AD theories with trivial Higgs branch, upon reduction on $S^1$ with the $\mathbb{Z}_N$ twist (where $\mathbb{Z}_N$ is a global symmetry of the given AD theory), result in the rank-0 3d $\mathcal{N}=4$ SCFTs, which have been a subject of recent studies. A generic AD theory, by the same logic, leads to a 3d $\mathcal{N}=4$ SCFT with zero-dimensional Coulomb branch (and suggests that there are a lot of them). Our construction therefore puts various empirical observations on the firm ground, such as, among other things, the match between the 4d VOA and the boundary VOA for some 3d rank-0 SCFTs previously observed in the literature. We end with an extensive list of promising open problems.}

\date{}

\maketitle

\tableofcontents

\section{Introduction}
Vertex Operator Algebras (VOAs) are abundant in Quantum Field Theory (QFT) in diverse dimensions, and appear whenever the problem has a preferred two-dimensional plane with the notion of holomorphy on it. This paper concerns VOAs in theories with extended supersymmetry (SUSY), found in the cohomology of a chosen nilpotent or equivariant supercharge. We aim to provide the setup that unifies several constructions of VOAs in SUSY QFT. By placing seemingly different research areas under the umbrella of a single construction, we uncover novel subtle connections and see new interesting problems.

We use the ideas akin to those of Nekrasov and Witten \cite{Nekrasov:2010ka} to reformulate the prototypical example of VOA in the higher-dimensional SUSY QFT -- the SCFT/VOA correspondence due to Beem-Lemos-Liendo-Peelaers-Rastelli-van Rees (BLLPRvR, or simply Beem et.al.) \cite{Beem:2013sza}. This reformulated SCFT/VOA correspondence will be our ``unifying device'' connecting it with other constructions. This is different from the more typical philosophy in modern physics, in which 6d theories play the role of ``unifying devices''. For us, it is the 4d SCFT with its 2d VOA.

According to the SCFT/VOA dictionary, each 4d $\cN=2$ superconformal field theory (SCFT) contains a supersymmetric sector described by a 2d VOA. It had been long suspected that an alternative formulation of the SCFT/VOA in terms of the Omega-deformation \cite{Nekrasov:2002qd,Nekrasov:2003rj,Nekrasov:2010ka} of the 4d theory is possible. Indeed, by 2019 such a formulation had been found. As it turns out, one has to start with the holomorphic-topological twist of the 4d $\cN=2$ SCFT \cite{Kapustin:2006hi} defined on
\begin{equation}
\Sigma\times C,
\end{equation}
where the Riemann surface $\Sigma$ is viewed as the holomorphic direction, and the two-dimensional surface $C$ is the topological direction. One then introduces the equivariant (or Omega) deformation along the topological plane $C$ \cite{Oh:2019bgz,Jeong:2019pzg,butson-unp} (see also \cite{Butson:2020coe,Butson:2020mmu}), assuming that $C\cong \R^2$ has an $S^1$-invariant metric. Without the deformation, observables in the HT-twisted theory form a Poisson Vertex Algebra \cite{Oh:2019mcg}. The deformation localizes them to the origin of $C$ and induces singular OPE along $\Sigma$.\footnote{Such reformulation of the six-dimensional construction \cite{Beem:2014kka} also exists \cite{Bobev:2020vhe}.} It is this Omega-deformation approach to the SCFT/VOA, together with $C$ having the cigar geometry, which will play the role of ``unifying device'' for us. Perhaps it sounds simple, and that is because it is. The consequences, however, are quite interesting and nontrivial. So what kind of constructions do we relate it with?

One closely related construction was introduced by Costello and Gaiotto in 3d $\cN=4$ theories \cite{Gaiotto:2016wcv,Costello:2018fnz}. They set out to define holomorphic boundary conditions in the topologically twisted versions of these theories. There are two fully topological twists: (\emph{i}) 3d B or C or Blau-Thompson or Rozansky-Witten twist \cite{Blau:1996bx,Rozansky:1996bq}, (\emph{ii}) and 3d A or H twist \cite{Kapustin:2010ag}, which is the reduction of Donaldson-Witten twist in 4d \cite{Witten:1988ze}. The construction of \cite{Costello:2018fnz} starts with some $\cN=(0,4)$ preserving boundary condition, which is not compatible with the topological twists. Then the boundary deformation is introduced that makes the boundary condition compatible with either the A or the B twist. This construction produces VOAs similar to those found in the Beem et.al. construction in 4d. The resemblance is close enough to suspect that they are directly related.

Another related construction is that of chiral algebras in 2d $\cN=(0,2)$ theories. Of course at the CFT point, 2d theories are controlled by chiral algebras. But even away from the CFT point, one can study the $\overline{Q}_+$ cohomology, which  has the structure of a VOA \cite{Witten:1993jg,Silverstein:1994ih} and is RG-invariant \cite{Dedushenko:2015opz}. It was noticed in \cite{Dedushenko:2017osi} that sometimes, the chiral algebra of such a theory coincides with the one in the SCFT/VOA correspondence. Though temping to dismiss as a coincidence, we find that it is not accidental, and in fact there is a direct link that connects the two.

Another line of developments we would like to connect with concerns the observations from \cite{Dedushenko:2018bpp,Fredrickson:2017yka,fredrickson2017s1fixed}. It turns out that 4d $\cN=2$ SCFTs, after the twisted supersymmetric $S^1$ reduction and the topological 3d A twist, may lead to interesting 3d TQFTs that capture the module category of the same VOA that appears in the SCFT/VOA correspondence. If in the parent 4d theory all Coulomb branch operators (i.e., coordinates on the Coulomb branch) have purely fractional conformal dimensions, and if in addition there is no Higgs branch, the 3d TQFT is especially nice, connected to a non-unitary modular tensor category (MTC). Roughly speaking, this happens for the following reason. In the presence of fractional Coulomb branch spectrum, the theory has a certain discrete flavor symmetry $\Z_N$. This is the case, for example, for most of the Argyres-Douglas theories \cite{Argyres:1995jj,Argyres:1995xn,Bonelli:2011aa,Xie:2012hs,Xie:2013jc}. We turn on the holonomy $(1\mod N)\in\Z_N$ around the $S^1$ durign the dimensional reduction. This holonomy, at least in the theories with purely fractional Coulomb spectrum, completely lifts the Coulomb branch. The geometric target simply degenerates to a discrete (and finite) collection of points. Some of these observations were based on explicit computations in \cite{Dedushenko:2018bpp}, and looked quite mysterious. Our unification construction will demystify this as well.

Furthermore, since the circle reduction of this sort naturally produces 3d $\cN=4$ theories without Higgs and Coulomb branches, this connects to yet another modern area of research. Namely, the 3d $\cN=4$ SCFTs without moduli spaces of vacua. They have recently received some attention in \cite{Gang:2018huc,Gang:2021hrd,Gang:2023rei,Ferrari:2023fez}, and were called the ``rank-zero'' theories. This terminology may be a little surprising to the experts on 4d $\cN=2$ SCFT, since in that case rank-zero means that the Coulomb branch is zero-dimensional, and it is believed that there are no interacting rank-zero theories \cite{Argyres:2020nrr}. The world of 3d $\cN=4$ SCFTs, however, is much richer, and contains a plethora of such examples. In particular, \cite{Gang:2018huc,Gang:2021hrd,Gang:2023rei} propose an infinite collection of rank-zero theories. We will also show that the Argyres-Douglas theories of type $(A_{k-1}, A_{M-1})$, with $\gcd(k,M)=1$, give rank-zero 3d $\cN=4$ SCFTs under the $\Z_{M+k}$-twisted circle reduction. The whole classification is not pursued here, and presumably involves many more examples.

For a brief road map of all the topics involved in our discussion, see Figure \ref{fig:unify}:
\begin{figure}[h]
	\centering
	\includegraphics[scale=0.8]{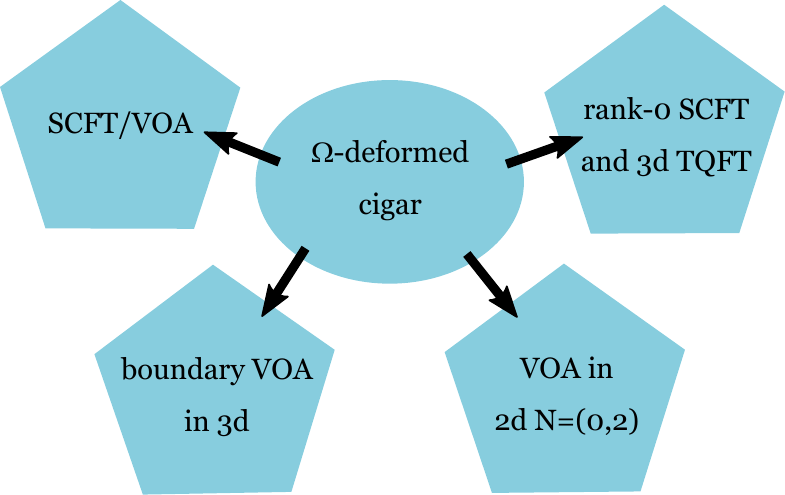}
	\caption{The areas where Vertex Algebras play role, and which are unified by our construction relying on the Omega-deformed cigar.\label{fig:unify}}
\end{figure}\\
So, as already mentioned earlier, our approach is based on applying the holomorphic-topological (HT) twist and the Omega-deformation to the 4d $\cN=2$ SCFT. We first put the undeformed 4d HT theory on $\Sigma\times C$, where $\Sigma$ is the holomorphic direction and $C$ is the topological direction having the shape of cigar. After degenerating the cigar $C$ into the half-line, we obtain a 3d $\cN=4$ theory with the canonical $(0,4)$ boundary condition denoted $H$ throughout this paper, see Figure \ref{fig:intro-cigar}:
\begin{figure}[h]
	\centering
	\includegraphics[scale=0.9]{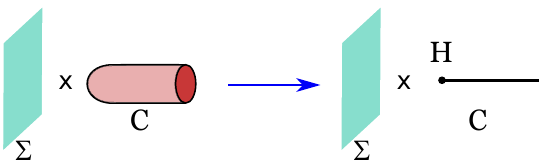}
	\caption{Degenerating the topologically twisted cigar engineers the boundary conditions $H$.\label{fig:intro-cigar}}
\end{figure}\\
We then turn on the Omega-deformation along $C$ with the parameter $\varepsilon$. This version of the Omega-deformation was constructed in \cite{Oh:2019bgz,Jeong:2019pzg}, in particular, \cite{Oh:2019bgz} showed that it looks like the Omega-deformed B-model \cite{Yagi:2014toa,Luo:2014sva} on $C$.

An important step is to realize, as in Nekrasov-Witten \cite{Nekrasov:2010ka}, that the Omega-deformation can be removed in the flat (cylindrical) region of the cigar via a simple field redefinition. It cannot be removed around the curved cap, however, where it is a non-trivial deformation. Then after degenerating $C$ into the half-line once again, we obtain the same undeformed 3d theory, subject to the \emph{deformed} boundary conditions $H_\varepsilon$. One of the important technical steps in this work is to obtain $H_\varepsilon$, and explicitly show that it coincides with the Costello-Gaiotto deformation from \cite{Costello:2018fnz}.

In Section \ref{sec:HTtwist} we start by reviewing the HT twist, the formulation of the 4d theory as a 2d model on $C$ with infinitely many fields, and the deformation of the flat $\R^2$ into the cigar. In Section \ref{sec:cig-bdy} we derive the $(0,4)$ boundary conditions $H$, and show how the Omega-deformation corresponds to their deformation to $H_\varepsilon$. We also study some properties of these boundary conditions and explain relation between different supercharges. In Section \ref{sec:DeformedBC}, after noting that $H_\varepsilon$ is the Costello-Gaiotto deformation of $H$, we elucidate various aspects of such deformations of the $(0,4)$ boundary conditions, including the explicit choice of the deforming boundary operator. In Section \ref{sec:tqft} we discuss the setup of 3d TQFT coupled to the holomorphic boundary conditions $H_\varepsilon$. This is where we discuss the $\Z_N$-twisted circle reduction from 4d, explain how it leads to the rank-zero theories, and make a conjecture about the precise rank-zero theory in the IR. In Section \ref{sec:2d02} we introduce a second boundary, chosen to be topological in the TQFT. We explain how such an interval reduction leads to the effective 2d $\cN=(0,2)$ model that knows a great deal of information about the boundary VOA and possibly extends it. We discuss Lagrangian examples, and mention the possibility of direct reduction on $C$ from 4d to 2d, and on $S^2$ \cite{Cecotti:2015lab}. Finally, in the Outlook Section, we collect many open problems posed as questions for the future throughout the paper.

\emph{Note added:} while working on this draft, we became aware of the recent paper \cite{Nawata:2023aoq}, which has a minor overlap with some of our ideas. Also \cite{Ferrari:2023fez} appeared, which may have a negligible overlap with our Section \ref{sec:DeformedBC}.
\subsection*{Acknowledgements} Some of the core ideas behind this work were the topic of my conversations with E.~Witten around 2014-2015. Back then I abandoned this project. Over the years, many technical ingredients of the story were developed by other people, which includes the Omega-deformation approach to the SCFT/VOA correspondence \cite{Oh:2019bgz,Jeong:2019pzg}, and the boundary VOA in 3d $\cN=4$ \cite{Costello:2018fnz} (\cite{Costello:2020ndc} is also relevant). Eventually, I decided to complete this work. Discussions with M.~Litvinov, who participated in the early part of the current phase of this project, played role in my decision. In the current phase of the project, I also enjoyed discussing this work with: C.~Beem, A.~Ferrari, N.~Nekrasov, L.~Rastelli, B.~Rayhaun, W.~Yan. These results were reported at the Pollica workshop on September 13, 2023, and also at the Berkeley mathematical physics seminar on November 27, 2023.

\section{Holomorphic-topological twist in 4d}\label{sec:HTtwist}
Consider the holomorphic-topological (HT) twist of a 4d $\cN=2$ SCFT, which was described in \cite{Kapustin:2006hi}. It allows to put SCFT on a four-manifold $M_4 = \Sigma\times C$, where $\Sigma$ and $C$ are two-dimensional surfaces. We view $\Sigma$ as a complex Riemann surface. It can be anything, but for our purposes it is mostly $\C$ or $\mathbb{T}^2$. This is the \emph{holomorphic} direction. We view $C$ as the \emph{topological} direction, which for us will mostly be the 2d cigar geometry, though other options like $S^2$ may appear as well. Throughout this paper, we loosely refer to $\Sigma$ as the holomorphic or chiral algebra plane, and to $C$ -- as the topological plane.

Let us fix conventions, in which the chiral $SU(2)_R$-doublet of supercharges is denoted $Q^i_\alpha$, where $i=1,2$ is an $SU(2)_R$ index, and the anti-chiral doublet is $\tQ_{i\dot\alpha}$:
\begin{equation}
\{Q^i_\alpha, \tQ_{j\dot\alpha}\}=\delta^i{}_j P_{\alpha\dot\alpha}.
\end{equation}
The $U(1)_r$ charge of $Q^i_\alpha$ is fixed to be $+\frac12$, and that of $\tQ_{i\dot\alpha}$ is $-\frac12$. The components $Q^i_\pm$ have spin $\pm\frac12$ under the chiral $\cM_+{}^+$, and $\tQ_{i\dot{\pm}}$ have spin $\pm\frac12$ under the anti-chiral $\cM^{\dot +}{}_{\dot +}$. These conventions agree with \cite{Beem:2013sza}, in particular, we also fix the rotations tangent to the $\Sigma$ and $C$ to be generated by:
\begin{equation}
\cM_\Sigma = \cM_+{}^+ + \cM^{\dot +}{}_{\dot +},\quad \cM_C = \cM_+{}^+ - \cM^{\dot +}{}_{\dot +}.
\end{equation}

The 4d HT twist uses in a non-trivial way both the $SU(2)_R$ and $U(1)_r$ R-symmetry groups, which is why it applies to SCFTs. The Lorentz group of $\Sigma$ is twisted by the maximal torus $U(1)_R\subset SU(2)_R$, and the Lorentz group of $C$ is twisted by $U(1)_r$. If we twisted both $\Sigma$ and $C$ by $U(1)_R$, this would be just a specialization of the Donaldson-Witten twist \cite{Witten:1988ze} to $M_4=\Sigma\times C$. Other choices would simply lead to the holomorphic twist \cite{Johansen:1994aw,Losev:1995cr,Losev:1996up,NikiThesis,Costello2011NotesOS,Budzik:2022mpd}.

Reduction of a 4d $\cN=2$ theory on $C$ with the topological twist \cite{Bershadsky:1995vm} by $U(1)_r$ preserves the following supercharges that are scalars along $C$:
\begin{equation}
\label{alg04}
Q^i_-,\quad \tQ_{i\dot{-}}.
\end{equation}
They form 2d $\cN=(0,4)$ SUSY on $\Sigma$. (Different twists result in 2d $\cN=(0,2)$ \cite{Cecotti:2015lab}.) 

Reduction of a 4d $\cN=2$ theory on $\Sigma$ with the $U(1)_R\subset SU(2)_R$ twist preserves the following supercharges that are scalars along $\Sigma$:
\begin{equation}
\label{alg22}
Q^1_-,\quad Q^2_+,\quad \tQ_{1\dot{+}},\quad \tQ_{2\dot{-}},
\end{equation}
giving $\cN=(2,2)$ supersymmetry on $C$. The overlap of \eqref{alg04} with \eqref{alg22} is $(Q^1_-, \tQ_{2\dot{-}})$, which are the two scalar supercharges in the 4d HT twist \cite{Kapustin:2006hi}. The standard notation for the 2d $\cN=(2,2)$ supercharges is $(Q_\pm, \overline{Q}_\pm)$, where the subscript indicates the chirality along $C$. The identification with \eqref{alg22} is:
\begin{equation}
\label{denote22}
Q_- = \tQ_{1\dot{+}},\quad Q_+ = Q^2_+,\quad \overline{Q}_- = Q^1_-,\quad \overline{Q}_+ = \tQ_{2\dot{-}}.
\end{equation}
From the perspective of this $\cN=(2,2)$ SUSY on $C$, a further twist of $C$ by $U(1)_R$ is like the 2d A-twist, while the twist of $C$ by $U(1)_r$ is the 2d B-twist ($U(1)_r$ is the axial R-symmetry). So once again, in the 4d HT twist, we perform the B-twist along $C$. According to \cite{Kapustin:2006hi}, the generic HT supercharge is given by the generic linear combination of the two scalar supercharges:
\begin{equation}
Q^1_- + t \tQ_{2\dot{-}}.
\end{equation}
It leads to the HT-twisted theory that is independent of $t$, as long as $t\neq 0,\infty$ (the theory is invariant under $\C^\times_r$, the complexification of $U(1)_r$, which allows to rescale $t$). We only consider the generic case, so for convenience pick $t=1$. Note that this HT supercharge, in the $\cN=(2,2)$ language, is equal to the standard B-model BRST differential on $C$:
\begin{equation}
\cQ_B = \overline{Q}_+ + \overline{Q}_- = Q^1_- + \tQ_{2\dot{-}},
\end{equation}
which is also the differential whose cohomology consists of the Schur operators in 4d \cite{Beem:2013sza,Oh:2019mcg}.

A theory with eight supercharges on $\Sigma\times C$ can be formulated as a 2d theory (with infinitely many fields) with four supercharges either on $\Sigma$ or on $C$. Such lower-dimensional formulations are well-known in other examples \cite{Gaiotto:2008sa,Bullimore:2016hdc,Bullimore:2016nji,Bullimore:2021auw,Ferrari:2022opl,Ferrari:2019cni}, and one can have at most half of SUSY manifest in such a description. Indeed, the full SUSY algebra in 4d involves translation in all four directions, while a 2d description should only involve two independent translations. Thus, just like Lorenz-invariant defects preserve at most half of SUSY, the lower-dimensional reformulations do the same.

In fact, one can come up with two such formulations compatible with the 4d HT twist that we pursue. One is a 2d $\cN=(2,2)$ theory on $C$ with the supercharges \eqref{alg22}; another is a 2d $\cN=(0,4)$ on $\Sigma$ with the supercharges \eqref{alg04}. We do not use the $\cN=(0,4)$ formulation here, however, we do use the $\cN=(2,2)$ one in a crucial way. It has been constructed (for Lagrangian theories) in \cite{Oh:2019bgz,Costello:2018txb}. Let us now describe the 2d $\cN=(2,2)$ construction in greater detail. As usual, it involves the infinite-dimensional gauge group:
\begin{equation}
\cG = {\rm Maps}(\Sigma, G),
\end{equation}
where $G$ is the 4d gauge group. One ingredient is the ${\rm ad}(\cG)$-valued $\cN=(2,2)$ vector multiplet $\cV$ on $C$, which gauges $\cG$ in the two-dimensional sense. This does not reproduce the full gauge kinetic term in 4d: One has to also include an ${\rm ad}(\cG)$-valued $\cN=(2,2)$ chiral multipelt $\cA_{\bar z}$, whose lowest component $A_{\bar z}$ is the antiholomorphic gauge connection along $\Sigma$. This multiplet does not quite transform in the adjoint representation of $\cG$, however. Just like it was observed in the 4d/3d case of \cite{Gaiotto:2008sa}, it transforms in the ``affine deformation'' of the adjoint of $\cG$, as instructed by the 4d gauge symmetry:
\begin{equation}
\label{aff_def}
\delta_u A_{\bar z} = -D_{\bar z}u = i[u,A_{\bar z}] - \partial_{\bar z}u,
\end{equation}
where $\delta_u$ denotes the infinitesimal $\cG$-gauge transformation with parameter $u$. According to the general principles, the gauge-covariant derivative of some field $\phi$ is defined as:
\begin{equation}
\cD_\mu \phi = \partial_\mu \phi + \delta_{A_\mu} \phi.
\end{equation}
where again $\delta_{\dots}$ denotes the infinitesimal $\cG$-gauge transformation.
In particular, the $\cG$-covariant derivative of $A_{\bar z}$ along $C$ (here $j=1,2$) is:
\begin{equation}
\cD_j A_{\bar z} = \partial_j A_{\bar z} + i[A_j, A_{\bar z}] - \partial_{\bar z} A_j = F_{j\bar{z}}.
\end{equation}
That is how we recover the mixed components of the gauge field strength, and the $g^{ij}F_{i z}F_{j\bar{z}}$ part of the kinetic term. Finally, the $F_{z\bar z}^2$ contribution to the 4d kinetic term looks like potential from the 2d viewpoint. More precisely, it is the D-term potential generated by the gauge transformation \eqref{aff_def}, as can be seen from the coupling between $A_{\bar{z}}$ and the auxiliary field $D$ living in $\cV$:
\begin{equation}
\int_\Sigma \dd z\, \dd\bar{z} \Tr\left(A_z \delta_D A_{\bar z} + A_{\bar z} \delta_D A_z\right) = \int_\Sigma \dd z\, \dd\bar{z}\,\Tr\, F_{\bar{z}z}D.
\end{equation}

To account for the 4d hypers valued in $\cR\oplus \bar{\cR}$, we include the $\cN=(2,2)$ chiral multiplets $\cX$ and $\cY$ that take their values in
\begin{equation}
{\rm Maps}(\Sigma,\cR) \quad \text{and}\quad {\rm Maps}(\Sigma,\bar\cR),
\end{equation}
respectively. The superpotential is
\begin{equation}
\label{W4d2d}
\cW = \int_\Sigma \cY\, \bar\partial_A \cX,
\end{equation}
where $\bar\partial_A = \dd\bar{z} (\partial_{\bar z} + A_{\bar z})$, and remember that the twist makes $\cY\cX$ into a $(1,0)$-form on $\Sigma$. The 2d kinetic term on $C$ for $\cX$ and $\cY$, as well as the potential $|\bar\partial_A \cX|^2+|\bar\partial_A \cY|^2$ originating from $\cW$, combine into the 4d kinetic term for the hypermultiplets. Additionally, the superpotential also generates the quartic D-term potential for the hypers.

Now recall \cite{Kapustin:2006hi} that the supercharge of the HT twist is the B-model supercharge $\cQ_B$ on $C$ (indeed, we perform the B-twist). It was further argued in \cite{Oh:2019bgz} that the Beem et. al. supercharge \cite{Beem:2013sza} can be identified with the Omega-deformed version of the B-model supercharge on $C$. Namely, assume the topological surface $C$ has a $U(1)$-invariant metric, with the $U(1)_\hbar$ isometry generated by $V$. There exists a deformation of the B-model supercharge, called the Omega-deformation \cite{Nekrasov:2002qd,Nekrasov:2003rj,Nekrasov:2010ka,Yagi:2014toa} and described below, which obeys:
\begin{equation}
\label{Omega-defB}
(\cQ_B)^2 = \hbar \cL_V.
\end{equation}
SUSY localization with respect to $\cQ_B$ restricts us to the fixed point(s) of $V$ along $C$. The residual degrees of freedom that remain dynamical on the localization locus are captured by a vertex operator algebra (VOA) along $\Sigma$. This is manifest since in the HT twist, $\Sigma$ is the holomorphic direction, hence the localization must produce something meromorphic along $\Sigma$. In the absence of Omega-deformation, this something is known to be a Poisson vertex algebra \cite{Oh:2019mcg} (which is commutative thanks to the presence of the topological direction $C$). The Omega-deformation quantizes it into a vertex operator algebra. In the Lagrangian case, it is determined from the localization of the Omega-deformed B-model on $C=\C^2$ (viewed as an infinite cigar) \cite{Costello:2018txb,Oh:2019bgz}. The residual integral over the localization locus is given by:
\begin{equation}
\label{LocAns}
\int_\cL {\rm Vol}_{\cL}\, e^{-\frac{i}{\hbar}\cW\big|_{0\in C}}.
\end{equation}
Here $\cL$ is the space of all gradient flows with respect to $\Re(\cW/\hbar)$ that end at $\cL_\infty\subset {\rm Crit}(\cW)$, where $\cL_\infty$ is itself a Lagrangian submanifold of the critical locus. One can think of $\cL$ as parameterizing the $r=0$ value of the gradient flow, while the endpoint corresponds to $r=\infty$. This answer has double meaning: (1) on the one hand, it simply follows from the Omega-deformed B-model localization on the infintie cigar \cite{Costello:2018txb}; (2) on the other hand, the same $\cL$ appears in the analytical continuation problem \cite{Witten:2010zr}, where we are simply trying to give mathematical meaning to the (path) integral \eqref{LocAns}. Hence \eqref{LocAns} indeed describes a two-dimensional QFT with the action $\cW$ given in \eqref{W4d2d}. It is a well known holomorphic 2d QFT called the \emph{gauged beta-gamma system} (or gauged symplectic boson), whose VOA --- the BRST reduction of the beta-gamma VOA --- is the answer for Lagrangian 4d SCFTs \cite{Beem:2013sza}.

While such an explicit picture of the cigar reduction is only available in the Lagrangian case, we assume that the general philosophy holds for non-Lagrangian theories as well. Any sort of twist, when performed in the flat space, is a trivial operation. It consists of relabeling fields and choosing the supercharge, whose cohomology we study. The same holds true for the HT twist. The case of Omega-deformation is slightly more subtle. In non-conformal theories, rotations do not appear in the SUSY algebra, and so to have something like \eqref{Omega-defB} as a symmetry, one really has to deform the theory \cite{Nekrasov:2002qd,Nekrasov:2010ka}.

In SCFT, however, rotations do appear in the superconformal algebra. Thus by using the ``$Q+S$'' type supercharge in flat space, as in the construction of \cite{Beem:2013sza}, one can emulate the Omega-deformation, namely have the supersymmetry that squares to the rotation. It was argued in \cite{Oh:2019bgz} that this not only emulates the Omega-deformation, but is in fact equivalent to it by a field redefinition. Their argument in 2d $\cN=(2,2)$ SCFT goes as follows: Consider the twisted and Omega-deformed theory on $\R^2$ and map it by the usual conformal map to $S^1\times \R$. On the latter, the rotation becomes the $S^1$ translation, and the Omega-deformation is removable by the field redefinition \cite{Nekrasov:2010ka}. In fact, precisely this removability is crucial in the current paper: we show it below both in the A and B model cases (the case of \cite{Nekrasov:2010ka} only covers the A-model). Thus after the field redefinition, we have an ordinary undeformed SCFT on $S^1\times \R$ with a choice of Poincare supercharge that squares to the translation along the $S^1$. Performing inverse conformal mapping back to $\R^2$, we recover the undeformed 2d $\cN=(2,2)$ SCFT there, but now with the ``$Q+S$'' type supercharges that squared to the rotation of $\R^2$. This argument clearly applies to our B-model realizing the HT twist, and shows that its Omega-deformation is equivalent to the chiral algebra construction of \cite{Beem:2013sza}.

B-twisted (or A-twisted) Omega-deformed theory allows only those deformations of the metric of $C$ that preserve the $U(1)$ isometry $V$. One such deformation we use is ``folding'' $\R^2$ into a cigar $C$. This deformation is $\cQ_B$-exact, which is the meaning of the adjective ``topological'' in front of the name ``twist''. Again, this can be explicitly checked in the Lagrangian case. Triviality of the metric deformation is equivalent to $\cQ_B$-exactness of the corresponding component of the (topologically twisted) 2d stress-enery tensor. Suppose the 4d stress-energy tensor is $T_{\mu\nu}$, with components $T_{ij}$ along $\Sigma$, $T_{ab}$ along $C$, and the mixed components $T_{ia}$. The 2d stress-energy tensor of the 2d $\cN=(2,2)$ theory is:
\begin{equation}
\theta_{ab} = \int_\Sigma T_{ab}\, {\rm Vol}_\Sigma.
\end{equation}
This tensor is conserved in 2d by adding a total divergence under the integral:
\begin{equation}
\partial^a \theta_{ab} = \int_\Sigma (\partial^a T_{ab}+ \partial^i T_{ib})\, {\rm Vol}_\Sigma=0.
\end{equation}
In the same way, if $J^r_\mu$ is the conserved 4d current for $U(1)_r$, we can define the conserved integrated current in 2d:
\begin{equation}
j_a = \int_\Sigma J^r_{a}\, {\rm Vol}_\Sigma.
\end{equation}
Under the B-twist in 2d, $\theta_{ab}$ is replaced by $\tilde{\theta}_{ab}$:
\begin{align}
\tilde{\theta}_{ww} &= \theta_{ww} - i\partial_w j_w,\cr
\tilde{\theta}_{w\bar{w}} &= \theta_{w\bar{w}} - i\partial_w j_{\bar w},\cr
\tilde{\theta}_{\bar{w}\bar{w}} &= \theta_{\bar{w}\bar{w}} + i\partial_{\bar w} j_{\bar w}.
\end{align}
The main property of the topological twist is that this $\tilde{\theta}_{ab}$ is $\cQ_B$-exact. In the presence of the Omega-deformation, only certain components of $\tilde{\theta}_{ab}$ remain $\cQ_B$-exact.
For the ``folding'' deformation, we deform the metric component $g_{\varphi\varphi}$ (here $\varphi$ is the angular coordinate on $\R^2$), so the corresponding stress-energy component must be exact:
\begin{equation}
\label{Q-exactness}
\tilde{\theta}_{\varphi\varphi} = \{\cQ_B, \dots\}.
\end{equation}
This is a statement about supercharges acting on conserved currents and supercurrents. It is completely encoded in the supercurrent multiplet and does not require any concrete realization of the theory. In particular, we start with a 4d $\cN=2$ SCFT, whose superconformal stress-energy multiplet (containing stress-energy tensor, supercurrents, R-symmetry currents, and additional operators \cite{Sohnius:1978pk}) is unique. Ultimately, this multiplet implies the relation \eqref{Q-exactness} in the 2d formulation. Hence the ideas described here hold equally well for general non-Lagrangian SCFTs. By the usual philosophy of QFT, we can perform the folding deformation generated by $\tilde{\theta}_{\varphi\varphi}$ in such a generality. The locality of QFT guarantees that in the end we get the same 4d SCFT placed on a different geometric background:
\begin{figure}[h]
	\centering
	\includegraphics[scale=1]{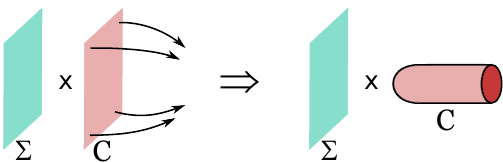}
	\caption{$\cQ_B$-exact deformation folding $C=\R^2$ into a cigar.\label{fig:fold}}
\end{figure}\\
In the process of folding, the topological twist induces a unit flux of the $U(1)_r$ symmetry through the curved region of the cigar (the tip):
\begin{equation}
\frac1{2\pi}\int_C F^{(r)} = 1.
\end{equation}
Asymptotically the cigar looks like a cylinder $S^1 \times \R$. Without the r-symmetry flux, we would end up with the NS spin structure on this $S^1$. Just like in the tt$^*$ geometry setup \cite{Cecotti:1991me}, the presence of flux creates a unit holonomy $e^{2\pi i}\in U(1)$ around the asymptotic circle. Writing $e^{2\pi i}$ makes it look trivial, however, the point is that a field of r-charge $r$ acquires an additional phase $e^{2\pi i r}$ around the $S^1$, which is nontrivial if $r$ is non-integral. If the 4d theory is Lagrangian (built from vector and (half-)hyper multiplets), then all the bosons have integer r-charges, and the spinors have r-charge $\pm\frac12$. Hence the phase $e^{2\pi i r}=-1$ converts the NS into the Ramond spin structure. Thus the 4d theory, far away from the tip of $C$, is dimensionally reduced on $S^1$ with the supersymmetric boundary conditions. In Lagrangian theories, this is the end of the story, as far as the $S^1$ reduction is concerned. Non-Lagrangian theories can have a richer spectrum of fractional r-charges, in which case $e^{2\pi i r}$ leads to interesting new phenomena that will be discussed in Section \ref{sec:3dTQFT}.

\section{From cigar to boundary conditions}\label{sec:cig-bdy}
Now let us assume $C$ has a concrete cigar geometry. We consider two versions of the cigar: An infinite $C\cong \R^2$, and a finite one $C\cong D^2$, which can be viewed as a disk with some boundary conditions fixed. In either case, the cigar metric is
\begin{equation}
\dd s^2 = \dd r^2 + \varrho(r)^2 \dd\varphi^2,
\end{equation}
where the radial coordinate $r$ runs from $0$ to $\infty$ in the infinite case, and from $0$ to $r_0$ in the finite case. Furthermore, in the infinite case we assume that the cigar approaches certain asymptotic radius:
\begin{equation}
\varrho(\infty)=R,
\end{equation}
whereas in the finite case, it takes on this value far enough from the tip:
\begin{equation}
\varrho(r)=R,\quad \text{for } a<r\leq r_0, \text{ where } a\ll r_0.
\end{equation}

A theory on $C$, in the region away from the tip, is effectively reduced on $S^1$ down to three dimensions. We will see that the three-dimensional theory naturally comes equipped with the twist, determined by the 4d twist and the presence or absence of the Omega-deformation. This setup is similar to the one by Nekrasov and Witten \cite{Nekrasov:2010ka}, the only distinction being that we work in a different twist (and different Omega-deformation) in this paper. In the case of infinite $C$, we thus obtain a 3d theory on the half-space, with the boundary conditions $H$ generated by the tip of cigar:
\begin{equation}
\Sigma \times C \Longrightarrow \Sigma \times \R_{r\geq 0},\quad \text{with the boundary conditions $H$ at } r=0.
\end{equation}
In the case of finite $C=D^2$, we obtain a 3d theory on the interval, with the second boundary condition $B$ originating from the 4d boundary conditions at $\Sigma\times \partial C$:
\begin{equation}
\Sigma \times C \Longrightarrow \Sigma \times I_{0< r < r_0},\quad \text{with the b.c. $H$ at } r=0 \text{ and $B$ at } r=r_0.
\end{equation}
\begin{figure}[h]
	\centering
	\includegraphics[scale=1]{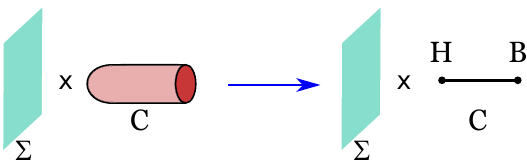}
\end{figure}\\
Let us first determine the boundary condition $H$ and its properties, as well as the 3d twist. This $H$ is canonical, up to two important choices: the choice of A or B twist along $C$, and whether it is Omega-deformed or not. The case of A-twist was studied in \cite{Nekrasov:2010ka}. Indeed, they worked in the Donaldson-Witten twist in 4d, which only uses $SU(2)_R$ and looks like the A-twist both along $\Sigma$ and $C$. We work with the B-twist along $C$ in the current paper.

\subsection{Empty twisted cigar}\label{sec:twisted_no_def}
First consider the cigar without Omega-deformation, which may be called ``empty''. In the case of A-twist, it is twisted by $U(1)_R\subset SU(2)_R$, and $H$ is a $(2,2)$ boundary condition, as follows from the identification of supercharges in \eqref{alg22}. In the case of B-twist, we twist by $U(1)_r$, and $H$ is a $(0,4)$ boundary condition as follows from \eqref{alg04}.

Reduction on the cigar does not constrain the values of fields at the tip, so long as they behave as scalars on $C$ in the chosen twist. Thus we expect $H$ to involve the Neumann boundary conditions on scalars (and their SUSY completion). To show this explicitly, consider a scalar $q$ on the $\Sigma\times C$ geometry. The equation of motion is
\begin{equation}
\Delta_\Sigma q + \partial_r^2 q + \frac{\rho'}{\rho} \partial_r q + \frac1{\rho^2} \partial_\varphi^2 q=0.
\end{equation}
We are interested in regular solutions in the limit of thin cigar, that is $\rho\to 0$. The last term in the above equation vanishes at $\partial_\varphi q=0$, which is the standard condition in dimensional reduction. The next to last term vanishes in the region where the radius of cigar is constant, $\rho'=0$. Near the tip of cigar, we find:
\begin{equation}
\frac{\rho'}{\rho} \sim \frac1{r}.
\end{equation}
Thus as $r\to0$, the regularity of solution implies
\begin{equation}
\partial_r q\big| =0.
\end{equation}
This semiclassical reasoning shows that $q$ obeys the Neumann boundary conditions in 3d. This is also enough to conclude that gauge fields obey the Neumann boundary conditions. Indeed, the components $A_i$ along $\Sigma$ behave like scalars on $C$, which, up to the standard gauge-theoretic technicalities, implies the Neumann boundary conditions,
\begin{equation}
F_{ir}\big|=0.
\end{equation}
Obviously, components $F_{ij}$ along $\Sigma$ are unconstrained by the cigar, as they should be along the Neumann boundary. The case of spinor fields is slightly more subtle to analyze.

Fortunately, we do not have to worry about the spinors since their boundary conditions, at least in our case, follow by SUSY. For the B-twist along $C$, the hypermultiplet scalars remain scalars, so they must obey Neumann boundary conditions (along with the gauge fields). The effective boundary should preserve $\cN=(0,4)$, as we know, so we obtain the $\cN=(0,4)$ Neumann boundary conditions on both hyper and vector multiplets. We will keep referring to these boundary conditions as $H$ below. In the A-twist, the compactification on $C$ preserves $\cN=(2,2)$, so we get some $(2,2)$ boundary conditions. This is part of the Nekrasov-Witten story, which we do not cover in this paper. 
To summarize:
\begin{align}
\text{2d B-twist} \Longrightarrow &(0,4) \text{ Neumann boundary conditions $H$ at } r=0.
\end{align}
To make contact with the readers familiar with the 2d A/B models, we also remark that after a further $S^1$ reduction to 2d, these boundary conditions have another known name:
\begin{align}
(0,4) \text{ boundary conditions } \leftrightarrow (B,B,B) \text{ branes}.
\end{align}

Let us briefly address the type of twist that the 3d theory comes equipped with. In 4d, we work in the HT twist, so the theory only depends on the complex structure along $\Sigma$, and does not couple to the metric along $C$. The holomorphic direction $\Sigma$ is unaffected by the reduction to 3d, while the topological direction $C$ degenerates to a line. The resulting 3d theory on $\Sigma\times\R$ is holomorphic along $\Sigma$ and topological along $\R$. This is the feature of the 3d HT twist \cite{Aganagic:2017tvx,Costello:2020ndc}, which only requires 3d $\cN=2$ SUSY. That is, the 4d HT (or Kapustin) twist reduces to the 3d HT twist applied to a 3d $\cN=4$ theory. The latter admits a further deformation to the full topological twist, which is known \cite{Garner:2022vds}, and will be relevant in the Section \ref{sec:BtwistOmega}. We summarize with the following schematic presentation, emphasizing the two equivalent descriptions in 4d, and their 3d reduction limit:
\begin{equation}
\label{scheme-no-Omega}
\boxed{\text{HT-twist on } \Sigma \times C}\quad  \vertarrowbox{\cong}{\boxed{\begin{matrix}\text{3d $\cN=4$, HT-twisted on } \Sigma\times\R_+,\\ \text{with $\cN=(0,4)$ boundary conditions }H\end{matrix}}}   \quad \boxed{\text{B-twist along } C}
\end{equation}
In particular, the B-model supercharge $\cQ_B$ on $C$, which determines the HT twist in 4d, reduces to the supercharge that determines the HT twist in 3d.

\subsubsection{Anomalies}\label{sec:anom}
The above characterization of boundary conditions assumes Lagrangian description. As said before, the general construction applies without such an assumption. The non-Lagrangian case is often challenging to study, but there is one thing that can be easily determined -- the anomaly polynomial of the boundary condition $H$. Indeed, the four-dimensional SCFT has a six-form anomaly polynomial \cite{Kuzenko:1999pi,Shapere:2008zf,Tachikawa:2015bga,Bah:2018gwc}:
\begin{equation}
\cI_6 = (a-c)(c_1(r)^3 - c_1(r)p_1(TM_4)) - 2(2a-c)c_1(r)c_2(R) + k c_1(r) n(f),
\end{equation}
where $c_1(r)$ is the first Chern class for the $U(1)_r$ and $c_2(R)$ is the second Chern class for the $SU(2)_R$ background gauge field, $p_1(TM_4)$ is the Pontryagin class of spacetime, and $n(f)$ is the appropriate Chern class of the background flavor gauge field. The coefficients $a$ and $c$ are the 4d Weyl anomalies and $k$ is the flavor central charge.

In a generic 4d SCFT, the anomaly polynomial of the boundary condition engineered by the cigar can be computed simply by pushing forward (integrating) $\cI_6$ along $C$ \cite{Alday:2009qq,Bah:2012dg} (though this might fail in the presence of accidental symmetries \cite{Bah:2017gph,Nardoni:2018dge}). For the A-twist, the flux of $U(1)_R\subset SU(2)_R$ through the tip of cigar breaks $SU(2)_R$ to $U(1)_R$, and it is the only reason why the pushforward of $\cI_6$ is nontrivial. Through the term $c_2(R)$, it picks up this flux:
\begin{equation}
\int_C c_2(R) = -\frac1{8\pi^2} \int_C \Tr (F_R^2) = -c_1(R),
\end{equation}
where now $c_1(R)$ is the first Chern class for the $U(1)_R$ gauge field. This implies the boundary anomaly polynomial:
\begin{equation}
P^{(2,2)}_\partial = 2(2a-c)c_1(A)c_1(V),
\end{equation}
where we identify $U(1)_r$ as the axial $U(1)_A$, and $U(1)_R$ -- as the vector $U(1)_V$ R-symmetry.

In the case of B-twist, it is the flux of $U(1)_r$ that contributes to the pushforward:\footnote{In our conventions, the supercharges have r-charge $\pm \frac12$. In the 3d limit, $U(1)_r$ is enhanced to $SU(2)_C$. With respect to the maximal torus $U(1)_C\subset SU(2)$, the fundamental weights of $SU(2)_C$ have charge $\pm1$. Thus the identification of $U(1)_r$ with $U(1)_C$ involves rescaling the generator by a factor of two. We take this into account below, when we rewrite $c_1(r)^2$ as $-4c_2(R_C)$.\label{foot:conv}}
\begin{equation}
\int_C c_1(r) = 1,
\end{equation}
implying the following boundary anomaly:
\begin{equation}
\label{anom_reduction}
P^{(0,4)}_\partial \equiv \cI_4 = -12(a-c) c_2(R_C) - (a-c)p_1(TM_4) - 2(2a-c)c_2(R_H) + k n(f).
\end{equation}
Here we have replaced $c_1(r)^2$ by $-4c_2(R_C)$, which captures the enhancement of $U(1)_r$ to $SU(2)_C$ in the 3d limit. Indeed, if the $SU(2)_C$ gauge field is only valued in the maximal torus $U(1)_C$, then $c_2=-\frac1{8\pi^2} [\Tr (F \sigma_3)^2] = -\frac1{4\pi^2} [F^2] = - c_1^2$ (see Footnote \ref{foot:conv} for the extra $4$).

For Lagrangian theories, these anomaly polynomials can be explicitly matched with the boundary anomalies of three-dimensional gauge theories, which are easy to compute. The anomaly polynomial of a 2d Weyl fermion (valued in the representation $\mathbf{R}$) is captured by the index density of the Dirac operator,
\begin{equation}
I_2 = [\hat{A}(R) {\rm ch}(F)]_4,\quad \hat{A}(R) = 1- \frac{p_1(TM_4)}{24} + \dots, \quad {\rm ch}(F)= \Tr_{\mathbf R} e^{F/(2\pi)},
\end{equation}
which leads to 
\begin{equation}
I_2=- \frac{p_1(TM_4)}{24}\dim_\C\mathbf{R} + \frac{c_1(F)^2}{2} - c_2(F).
\end{equation}
When the same Weyl fermion appears as a boundary mode of the 3d Dirac fermion, the anomaly polynomial is one half of such density, $\frac12 I_2$ \cite{Horava:1995qa,Horava:1996ma,Dimofte:2017tpi}. Note that this includes the gravitational anomaly  \cite{Alvarez-Gaume:1983ihn}. In particular, a 3d $\cN=2$ chiral multiplet with $(0,2)$ Dirichlet boundary conditions contributes $-\frac1{48}p_1$, for $(0,2)$ Neumann we get $\frac1{48}p_1$, a vector multiplet with the $(0,2)$ Dirichlet contributes $\frac1{48}p_1$, and a vector with the $(0,2)$ Neumann contributes $-\frac1{48}p_1$ to the boundary gravitational anomaly.

To address the B-twist, we write the boundary anomaly of the $\cN=(0,4)$ Neumann\footnote{As a reminder, a 3d $\cN=4$ vector multiplet consists of a 3d $\cN=2$ vector $\cV$ and a 3d $\cN=2$ adjoint-valued chiral $\Phi$. The $\cN=(0,4)$ Neumann boundary conditions \cite{Chung:2016pgt,Hanany:2018hlz} on the $\cN=4$ vector multiplet consist of the $(0,2)$ Neumann on $\cV$ and the $(0,2)$ Dirichlet on $\Phi$ \cite{Yoshida:2014ssa}. The $(0,4)$ Neumann boundary conditions on the hypermultiplet impose the $(0,2)$ Neumann boundary conditions on the constituent chiral multiplets.} boundary conditions:
\begin{equation}
\label{bndrAnom04}
P^{(0,4)}_\partial = \frac{{\rm Tr}_{\rm adj} (\mathbf{f}^2)}{8\pi^2} + \frac{|G|}{2} \frac{(\mathbf{r}_H-\mathbf{r}_C)^2}{8\pi^2} + \frac{|G|}{2} \frac{(\mathbf{r}_C + \mathbf{r}_H)^2}{8\pi^2} - {\rm Tr}_{\mathbf{R}} \frac12 \frac{(\mathbf{f} + \mathbf{f}_f - \mathbf{r}_C)^2}{8\pi^2} - \frac{2|G|-|\mathbf{R}|}{48}p_1(TM_4).
\end{equation}
Here $\mathbf{f}$ is the field strength for the gauge group, $\mathbf{f}_f$ is the same for the flavor group, $\mathbf{r}_C$ corresponds to the Cartan $U(1)_C\subset SU(2)_C$, and $\mathbf{r}_H$ -- to the Cartan $U(1)_H\subset SU(2)_H$. The space $\mathbf{R}$ is the quaternionic matter representation, in particular, for the full hypers it is given by $\cR\oplus \bar{\cR}$. The notation $h$ stands for the dual Coxeter number, $|G|$ is the real dimensions of $G$, and $|\mathbf{R}|$ is the complex dimension of $\mathbf{R}$.
One can observe that the $U(1)_r\times G^2$ anomaly in 4d ($G$ is the gauge group), i.e., the axial anomaly of $U(1)_r$, is mapped to the boundary gauge anomaly in 3d. The former should vanish for the B-twist to make sense, and the latter -- for the boundary condition to exist. Indeed, the boundary gauge anomaly in \eqref{bndrAnom04} vanishes if:
\begin{equation}
{\rm Tr}_{\rm adj} (\mathbf{f}^2) - \frac12 {\rm Tr}_{\mathbf{R}} (\mathbf{f}^2)=(2h-T_{\mathbf R}/2) \frac1{2h} {\rm Tr}_{\rm adj}(\mathbf{f}^2)=0,
\end{equation}
which is precisely the condition for cancellation of the axial anomaly of $U(1)_r$ in 4d, and also the condition for conformality due to the $\cN=2$ SUSY. After cancellations, and using the standard notations:
\begin{equation}
|G|=n_v, \qquad |\mathbf{R}|=2n_h,
\end{equation}
we obtain:
\begin{equation}
P^{(0,4)}_\partial = -\frac{n_v-n_h}{2}c_2(R_C) - \frac{n_v}{2} c_2(R_H) - \frac{n_v-n_h}{24}p_1(TM_4) - {\rm Tr}_{\mathbf{R}} \frac12 \mathbf{f}_f^2.
\end{equation}
The relation between $(n_v, n_h)$ and $(a,c)$ is well-known:
\begin{equation}
a-c=\frac{n_v-n_h}{24},\quad 2a-c = \frac{n_v}{4}.
\end{equation}
After such identifications, this anomaly polynomial precisely matches \eqref{anom_reduction}. Of course, this match by itself is not a novel result, just a sanity check.

What is important, though, is that in Lagrangian theories, $n_v$ is an integer and $n_h$ may be half-integral in the presence of half-hypermultiplets. One can shift coefficients in the boundary anomaly polynomial by adding boundary degrees of freedom, but the Lagrangian boundary degrees of freedom cannot change the quantization of $(n_v, n_h)$ entering $P^{(0,4)}_\partial$. For example, looking at the gravitational anomaly, its coefficient $-\frac{n_v-n_h}{24}=c-a$ is always proportional to $\frac1{48}$ at a Lagrangian boundary of a Lagrangian theory.\footnote{Here we are really talking about Lagrangian 3d $\cN=2$ theories with $(0,2)$ boundary conditions of the Dirichlet/Neumann type, and possible boundary multiplets, including chiral, Fermi and vector multiplets. If we allow $\cN=1$ or no SUSY, then the boundary gravitational anomaly may be proportional to $\frac1{96}$, which is the case for a chiral boundary condition of a 3d Majorana fermion. This only slightly refines the argument by a factor of 2, without affecting the conclusion, since there are still plenty of theories, for which $96(a-c)$ is fractional.} Thus $48(a-c)$ must be an integer. Of course there exist a lot of 4d $\cN=2$ SCFTs, whose $48(a-c)$ is fractional, implying that for such theories the corresponding boundary condition $H$ cannot be engineered in a Lagrangian way (even if the 4d theory admits $\cN=1$ Lagrangian).

One more remark is due. We are about to Omega-deform the boundary condition $H$. This will not affect the boundary anomaly, since the boundary anomaly is a discrete quantity, while the Omega-deformation is controlled by the continuous parameter $\varepsilon$.

\subsection{Twisted and Omega-deformed cigar}\label{sec:BtwistOmega}
Now let us explicitly describe the Omega-deformation of the 2d B-model as in \cite{Yagi:2014toa,Luo:2014sva,Nekrasov:2018pqq}. We will show that on the cigar $C$, the Omega-deformation can be removed far away from the tip by a simple field redefinition. This trick was used by Nekrasov and Witten in the 2d A-model case \cite{Nekrasov:2010ka}. It is essential for understanding the effect of Omega-deformation in the 3d limit, both on the 3d theory and on the boundary conditions $H$. After the field redefinition, we obtain the usual undeformed 4d theory far away from the tip, albeit with the different choice of supercharge. In particular, it no longer reduces to the supercharge of the 3d HT twist, rather, we get the fully topological 3d A-twist. So in the 3d limit, the Omega-deformation along $C$ does two things: (1) deforms the 3d HT twist to the fully topological 3d A-twist (realizing the deformation from \cite{Garner:2022vds}); (2) deforms the $(0,4)$ boundary conditions $H$ to the holomorphic boundary conditions compatible with the 3d A-twist (realizing the deformation from \cite{Costello:2018fnz}, which will be further discussed in Section \ref{sec:DeformedBC}).

\subsubsection{B-twisted and Omega-deformed}
As a warm-up, we start with the B-twisted Landau-Ginzburg model \cite{Yagi:2014toa,Nekrasov:2018pqq}. The fields of the model are: (1) complex scalars $x^i$ and their conjugates $\bar{x}^{\bar i}$ describing the sigma-model map; (2) worldsheet scalar $\eta^{\bar i}$ valued in the $(0,1)$ vector fields on the target; (3) worldsheet one-form $\rho^i_\mu$ valued in the $(1,0)$ vector fields on the target; (4) worldsheet two-form $\mu^{\bar i}_{z\bar{z}}$ valued in the $(0,1)$ vector fields on the target; (5) auxiliary worldsheet two-forms $F^i_{z\bar{z}}, \bar{F}_{z\bar{z}}^{\bar i}$ valued in the $(1,0)$ and $(0,1)$ vector fields on the target, respectively. With $V$ denoting the isometry of the target, the Omega-deformed SUSY is:
\begin{align}
\delta_\varepsilon x^i = \varepsilon \iota_V \rho^i,\quad \delta_\varepsilon \bar{x}^{\bar i} &= \eta^{\bar i},\cr
\delta_\varepsilon \rho^i = \dd x^i + \varepsilon \iota_V F^i,\quad \delta_\varepsilon \eta^{\bar i} &= \varepsilon \cL_V \bar{x}^{\bar i},\cr
\delta_\varepsilon F^i = \dd \rho^i,\quad \delta_\varepsilon\mu^{\bar i}=\bar{F}^{\bar i},\quad \delta_\varepsilon \bar{F}^{\bar i}&=\varepsilon\dd \iota_V \mu^{\bar i},
\end{align}
where the target is taken to be flat for simplicity. The action is normally defined as:
\begin{equation}
S = S_0 + S_W,
\end{equation}
where
\begin{align}
S_0&=\delta_\varepsilon \int (g_{i\bar{j}}\rho^i \star (\dd x^{\bar j} + \varepsilon \iota_V \bar{F}^{\bar j}) + g_{i\bar{j}}F^i \star \mu^{\bar j}),\\
S_W &= i\int (F^i \partial_i W + \frac12 \rho^i \rho^j \nabla_i\partial_j W + \delta_\varepsilon (\mu^{\bar i}\partial_{\bar i}\overline{W})).
\end{align}
The bosonic part of the action is
\begin{equation}
S_{\rm bos}=\int \left( g_{i\bar{j}} (\dd x^i + \varepsilon \iota_V F^i) \star (\dd x^{\bar j} + \varepsilon \iota_V \bar{F}^{\bar j}) + g_{i\bar{j}}F^i \star \bar{F}^{\bar j} + i F^i\partial_i W + i \bar{F}^{\bar i}\partial_{\bar i} \overline{W}  \right),
\end{equation}
which is actually valid in curved target too. Now consider this action on the flat cylinder with coordinates $r$ and $\varphi$, since the cigar away from the tip looks like one. We write $F^i =F^i \dd r\, \dd\varphi$ and obtain:
\begin{align}
\label{OmBver1}
S_{\rm bos}&=\int \dd r\, \dd\varphi \left( g_{i\bar{j}} (\partial_r x^i - \varepsilon F^i) (\partial_r x^{\bar j} - \varepsilon \bar{F}^{\bar j}) + g_{i\bar{j}} \partial_\varphi x^i \partial_\varphi x^{\bar j} + g_{i\bar{j}}F^i \bar{F}^{\bar j} + i F^i\partial_i W + i \bar{F}^{\bar i}\partial_{\bar i} \overline{W}  \right),\cr
&=\int \dd r\, \dd\varphi \Big( g_{i\bar{j}} \partial_r x^i \partial_r x^{\bar j} - \varepsilon (F_{\bar j}\partial_r x^{\bar j} + \bar{F}_i \partial_r x^i) + g_{i\bar{j}} \partial_\varphi x^i \partial_\varphi x^{\bar j} + (1+\varepsilon^2)g_{i\bar{j}}F^i \bar{F}^{\bar j}\cr &+ i F^i\partial_i W + i \bar{F}^{\bar i}\partial_{\bar i} \overline{W}  \Big).
\end{align}
By shifting $F$ and $\bar{F}$ to remove terms linear in $F, \bar{F}$, this turns into:
\begin{align}
\int \dd r\, \dd\varphi \Big( \frac{1}{1+\varepsilon^2} g_{i\bar{j}} \partial_r x^i \partial_r x^{\bar j}  + g_{i\bar{j}} \partial_\varphi x^i \partial_\varphi x^{\bar j} + (1+\varepsilon^2)g_{i\bar{j}}F^i \bar{F}^{\bar j}\cr + i F^i\partial_i W + i \bar{F}^{\bar i}\partial_{\bar i} \overline{W}  \Big)
+ \int \dd r\, \dd\varphi \frac{i\varepsilon}{1+\varepsilon^2} \partial_r (W+\overline{W})
\end{align}
The last term disappears, being a total derivative. By further rescaling $r$, metric $g_{i\bar{j}}$, auxiliary fields, and the superpotential,
\begin{equation}
r\mapsto \frac{r}{\sqrt{1+\varepsilon^2}},\quad g_{i\bar{j}}\mapsto \sqrt{1+\varepsilon^2} g_{i\bar{j}},\quad F^m \mapsto \frac{F^m}{\sqrt{1+\varepsilon^2}},\quad W\mapsto (1+\varepsilon^2)W.
\end{equation}
the deformation is fully removed from the bosonic action.

We can slightly simplify the above procedure. First note that we are free to modify the action by $\delta_\varepsilon$-exact terms as we please. In particular, we may drop the piece that contains $\varepsilon$ under the $\delta_\varepsilon$-variation, and define a slightly simpler kinetic action:
\begin{equation}
\label{SimpleS0}
S_0=\delta_\varepsilon \int (g_{i\bar{j}}\rho^i \star \dd x^{\bar j} + g_{i\bar{j}}F^i \star \mu^{\bar j}).
\end{equation}
In the case of constant metric, the fermionic part of this action has no $\varepsilon$ in it, which is one advantage of \eqref{SimpleS0}:
\begin{equation}
S_0^{\rm ferm} = \int (g_{i\bar{j}} \rho^i \star \dd \eta^{\bar j} + g_{i\bar{j}} \dd\rho^i \star \mu^{\bar j}).
\end{equation}
The bosonic action is:
\begin{align}
S_0^{\rm bos}&=\int \dd r\, \dd\varphi \left( g_{i\bar{j}} (\partial_r x^i - \varepsilon F^i) \partial_r x^{\bar j} + g_{i\bar{j}} \partial_\varphi x^i \partial_\varphi x^{\bar j} + g_{i\bar{j}}F^i \bar{F}^{\bar j} + i F^i\partial_i W + i \bar{F}^{\bar i}\partial_{\bar i} \overline{W}  \right).
\end{align}
It may look slightly asymmetric (with respect to $x^i \leftrightarrow \bar{x}^{\bar i}$) compared to the action \eqref{OmBver1}. However, the advantage here is that the $\varepsilon$-deformation is removed by the simple shift:
\begin{equation}
\label{Bshift}
\bar{F}^{\bar j} \mapsto \bar{F}^{\bar j} + \varepsilon \partial_r x^{\bar j}, 
\end{equation}
without any further rescalings. The shift only generates a total derivative term in the action:
\begin{equation}
S = S\big|_{\varepsilon=0} + i\varepsilon \int \dd r\, \dd\varphi\, \partial_r \overline{W}.
\end{equation}
These conclusions continue to hold for the curved target as well, we just need to remember to include $-\varepsilon(\iota_V \rho^k)\Gamma^i_{kj}\rho^j$ in $\delta_\varepsilon\rho^i$ and $-\varepsilon (\iota_V \rho^k)\Gamma^i_{kj} F^j$ in the definition of $\delta_\varepsilon F^i$. Thus indeed, we remove the $\Omega$-deformation on the cylinder by a simple field redefinition \eqref{Bshift}.

What if the cylinder is replaced by our cigar, i.e., one of its ends is capped off? The cigar metric, we remind, is
\begin{equation}
\dd s^2 = \dd r^2 + \varrho(r)^2 \dd\varphi^2.
\end{equation}
We are still using the action $S_0 + S_W$, with $S_0$ defined in \eqref{SimpleS0}. The same field redefinition \eqref{Bshift} results in the same effect of removing the $\varepsilon$-deformation almost completely, up to the derivative term:
\begin{equation}
S = S\big|_{\varepsilon=0} + i\varepsilon \int \varrho(r)^2\dd r\, \dd\varphi\, \partial_r \overline{W}.
\end{equation}
After the integration by parts, the term at infinity vanishes. Now, however, there is a contribution from the curved region near $r=0$ where $\varrho'\neq 0$ (recall that $\varrho(0)=0$):
\begin{equation}
S = S\big|_{\varepsilon=0} - 2i\varepsilon \int \varrho\varrho'\dd r\, \dd\varphi\, \overline{W}.
\end{equation}
In the limit of infinitely thin cigar, this leads to the boundary term in 1d:
\begin{equation}
-2\pi i \varepsilon R^2 \overline{W}\big|_{r=0},
\end{equation}
where $R=\varrho(+\infty)$.

\paragraph{$\Omega$-deformed B-twisted gauge theory.} Let us generalize this to B-twisted gauge theories constructed in \cite{Luo:2014sva}. The twisted vector multiplet contains bosonic one-forms $A, \sigma$, a bosonic auxiliary two-form $\mathbf{D}$, a fermionic one-form $\lambda$, fermionic two-forms $\zeta$ and $\alpha$. The SUSY is:
\begin{align}
\delta_\varepsilon A &= i\lambda,\quad \delta_\varepsilon\sigma=\lambda + \varepsilon \iota_V \zeta,\cr
\delta_\varepsilon\lambda &= -i\varepsilon \iota_V F_A + \varepsilon \dd_A \iota_V \sigma,\quad \delta_\varepsilon\zeta = i F_A +\dd_A \sigma - \sigma\wedge\sigma,\quad \delta_\varepsilon\alpha = \dd_A\sigma + \mathbf{D},\cr
\delta_\varepsilon \mathbf{D} &= \varepsilon \left(\dd_A\iota_V\alpha -[\iota_V\sigma,\alpha] -\dd_A\lambda -\lambda\wedge\sigma - \sigma\wedge\lambda -\dd_A\iota_V\zeta\right).
\end{align}
Here $\dd_A=\dd-iA$. The vector multiplet action is:
\begin{align}
S_V &= \delta_\varepsilon \int {\rm Tr} \left(\alpha\wedge\star(-\dd_A \sigma +\mathbf{D}+4\bar{\partial}_A\sigma) +\zeta\wedge \star (-i F_A + \dd_A\sigma + \sigma\wedge\sigma)\right)\cr
&= \int {\rm Tr} \Big( F_A \star F_A + \sigma \star \Delta\sigma + \frac{R}{2} \sigma \star\sigma - (\sigma \wedge\sigma)\star(\sigma\wedge\sigma)\cr &\qquad + \mathbf{D}'\star \mathbf{D}' +2\partial_A(\sigma\star \bar{\partial}_A\sigma) + 2\bar{\partial}_A(\sigma\star \partial_A\sigma) \cr 
&\qquad -2\alpha \star \dd_{A-i\sigma}(\lambda^{1,0} - \lambda^{0,1}) - 2\zeta\star \dd_{A+i\sigma}\lambda \cr
&\qquad -\varepsilon \alpha\star\dd_{A-i\sigma}\iota_V\alpha - \varepsilon\zeta \star \dd_{A+i\sigma} \iota_V \zeta -2\varepsilon \alpha\star \dd_A ((\iota_V\zeta)^{1,0} - (\iota_V\zeta)^{0,1}) \Big).
\end{align}
The bosonic action already has no $\varepsilon$ and requires no redefinitions. As for the fermionic part, after some massaging and integration by parts, the last two lines become:
\begin{equation}
-2\alpha\wedge\star \dd_{A-i\sigma} (\tilde\lambda^{1,0} -\tilde\lambda^{0,1}) - 2\zeta\wedge \star \dd_{A+i\sigma} (\tilde\lambda^{1,0} + \tilde\lambda^{0,1}),
\end{equation}
where
\begin{align}
\tilde\lambda^{1,0} &= \lambda^{1,0} + \frac{\varepsilon}{2}(\iota_V\alpha)^{1,0} + \frac{\varepsilon}{2}(\iota_V \zeta)^{1,0},\cr
\tilde\lambda^{0,1} &= \lambda^{0,1} - \frac{\varepsilon}{2}(\iota_V\alpha)^{0,1} + \frac{\varepsilon}{2}(\iota_V \zeta)^{0,1}.
\end{align}
or simply:
\begin{equation}
\label{VecFermRedef}
\tilde\lambda = \lambda + \frac{\varepsilon}{2} i\star (\iota_V \alpha) + \frac{\varepsilon}{2} \iota_V\zeta.
\end{equation}
This expression clearly gives the desired change of variables that removes the $\varepsilon$-deformation from the vector multiplet action. What happens if we are on the cigar rather than the cylinder? The only dangerous step is integration by parts. It involves the term
\begin{equation}
-\varepsilon\alpha\wedge\star \dd_{A+i\sigma}((\iota_V\zeta)^{1,0}-(\iota_V\zeta)^{0,1}) = \varepsilon \dd_{A+i\sigma}((\iota_V\zeta)^{1,0}-(\iota_V\zeta)^{0,1})\wedge \star\alpha,
\end{equation}
which after the integration by parts becomes:
\begin{equation}
\varepsilon ((\iota_V\zeta)^{1,0}-(\iota_V\zeta)^{0,1})\wedge \dd_{A+i\sigma}\star\alpha,
\end{equation}
and does not earn us any boundary terms. Further manipulations are also not affected by the curved metric of the cigar. We conclude that we do \emph{not} earn any boundary terms in the process of removing the $\varepsilon$-deformation of the vector multipelt action.

Next we look at the chiral multiplet. It has the same field content as in the ungauged LG example: $(x^m, \rho^i, \eta^{\bar i}, \mu^{\bar i}, F^i, \bar{F}^{\bar i})$. The SUSY is
\begin{align}
\delta_\varepsilon x^i &= \varepsilon\iota_V \rho^i, \quad \delta_\varepsilon \bar{x}^{\bar i} = \eta^{\bar i},\cr
\delta_\varepsilon \rho^i &= \dd_A x^i - \sigma x^i + \varepsilon \iota_V F^i,\quad \delta_\varepsilon\eta^{\bar i} = \varepsilon\left( \iota_V \dd_A \bar{x}^{\bar i} + \bar{x}^{\bar i}\iota_V \sigma \right),\quad \delta_\varepsilon \mu^{\bar i}=\bar{F}^{\bar i},\cr
\delta_\varepsilon F^i &= \dd_A\rho^i -\sigma \wedge \rho^i + \zeta x^i,\quad \delta_\varepsilon \bar{F}^{\bar i} = \varepsilon \left( \dd_A \iota_V \mu^{\bar i} + \mu^{\bar i}\iota_V\sigma \right).
\end{align}
The chiral multiplet action again consists of the $\delta_\varepsilon$-extact kinetic term $S_C$ and the $\delta_\varepsilon$-closed superpotential term $S_W$. The former is defined in \cite{Luo:2014sva} as the $\delta_\varepsilon$ variation of $\cV_\varepsilon$, where $\cV_\varepsilon$ is itself an $\varepsilon$ deformation of the $\varepsilon=0$ case. Like in the LG example above, we will rather use the undeformed $\cV_{\varepsilon=0}$. This is a $\delta_\varepsilon$-exact modification of the action, which does not affect the physical quantities. At the same time, it gives a version of the Omega-deformation that is removed by a simpler field redefinition on the cylinder. The action is:
\begin{align}
S_C &= \delta_\varepsilon \int \left( \rho\wedge \star (\dd_A\bar{x} - \bar{x}\sigma) - i x \bar{x}\alpha + F\wedge\star \mu  \right)\cr
&= \int \big( (\dd_A x^i - \sigma x^i + \varepsilon \iota_V F^i)\wedge \star (\dd_A\bar{x} - \bar{x}\sigma)\cr &- \rho\wedge \star (\dd_A\eta - \bar{x}\lambda - \eta\sigma - \bar{x}(\lambda + \varepsilon\iota_V\zeta)) - i \varepsilon \iota_V \rho \bar{x}\alpha - i x \eta\alpha\cr &- i x \bar{x} (\dd_A + \mathbf{D}) + F\wedge\star \bar{F} + (\dd_A\rho -\sigma \wedge \rho + \zeta x)\wedge\star \mu  \big).
\end{align}
The superpotential term:
\begin{align}
S_W = i\int \left( F \partial W + \bar{F}\overline{\partial W} + \frac12 \rho\wedge\rho \frac{\partial^2 W}{\partial x \partial x} +\eta \mu \frac{\partial^2 \overline{W}}{\partial \bar{x} \partial \bar{x}} \right) - \frac{i}{\varepsilon} \int_{\partial} W \dd\varphi.
\end{align}
The part of fermionic action inside $S_C$ involving $\varepsilon$ can be written as:
\begin{equation}
2\rho \wedge \star \bar{x}\left(\lambda + \frac{\varepsilon}{2}i\star (\iota_V\alpha) + \frac{\varepsilon}{2} \iota_V\zeta\right),
\end{equation}
thus the field redefinition \eqref{VecFermRedef} precisely removes the $\varepsilon$-dependent fermionic terms. The remaining bosonic term $\varepsilon (\iota_V F)\wedge \star (\dd_A \bar{x} - \bar{x}\sigma)$ is removed by shifting $\bar{F}$:
\begin{equation}
\label{shift_gauged}
\bar{F}^{\bar j} \mapsto \bar{F}^{\bar j} + \varepsilon (\partial_r \bar{x}^{\bar j} + \bar{x}^{\bar j} (iA_r - \sigma_r)).
\end{equation}
This shift affects the term $\bar{F}\overline{\partial W}$ in $S_W$. The shift by $\bar{x}^{\bar j} (iA_r - \sigma_r)$ in \eqref{shift_gauged} represents a gauge transformation of $\overline{W}$ and hence cancels. The remaining shift gives $\varepsilon \partial_r \bar{x}^{\bar j}\bar{\partial}_j \overline{W} = \varepsilon \partial_r \overline{W}$. In the case of cigar, we therefore obtain the same term as in the ungauged LG case:
\begin{equation}
i\varepsilon\int \partial_r \overline{W}\, \varrho(r)^2 \dd r\, \dd\varphi.
\end{equation}
We again integrate by parts, and after the reduction, find the boundary term:
\begin{equation}
-2\pi i \varepsilon R^2 \overline{W}\big|_{r=0}.
\end{equation}

Thus we see very explicitly that the effect of Omega-deformation on the boundary conditions $H$ is to deform them by the boundary term $\overline{W}$. The coefficient in front of the boundary deformation is of course proportional to $\varepsilon$. We will sometimes refer to the deformed boundary conditions as $H_\varepsilon$. Now we note that the supercharge is also affected by this deformation. The usual B-model supercharge,
\begin{equation}
\cQ_B = \overline{Q}_+ + \overline{Q}_-,
\end{equation}
upon the Omega-deformation, it is replaced by:
\begin{equation}
\cQ_B + \varepsilon \iota_V G,
\end{equation}
where $G_\mu$ is the one-form supercharge that is part of the B-twisted version of the $\cN=(2,2)$ Poincare SUSY. Here $V=\partial/\partial\varphi$, so $\iota_V G$ is the angular component of $G$. In the flat region of cigar, where $C$ looks like $\R\times S^1$, this $\iota_V G$ is simply the component of $G$ along $S^1$. It is given by the sum of the other two Poincare supercharges:
\begin{equation}
\iota_V G = Q_+ + Q_-.
\end{equation}
Thus the Omega-deformed supercharge in that region can be viewed as
\begin{equation}
\label{QBdefomred}
\overline{Q}_+ + \overline{Q}_- + \varepsilon(Q_+ + Q_-).
\end{equation}
This is just the special choice of the Poincare supercharge, which is why we were able to remove the Omega-deformation in this region by a field redefinition. From the viewpoint of the reduced theory on $\R_+$, the supercharge $\overline{Q}_+ + \overline{Q}_-$ is thus deformed to \eqref{QBdefomred}. This is how below, the 3d HT twist will get deformed to the fully topological A-twist.

\subsubsection{Application to the 4d gauge theory}
Now let us look at the 4d gauge theory subject to the holomorphic-topological (HT) twist with the Omega-deformation along the topological plane. We explained in Section \ref{sec:HTtwist} how to formulate the 4d theory as a 2d $\cN=(2,2)$ model on the topological plane with infinitely many fields. It was noted in \cite{Oh:2019bgz} that the 4d HT twist is equivalent to the B-twist of this 2d model. We have also reviewed how to Omega-deform the B-model \cite{Yagi:2014toa,Luo:2014sva,Nekrasov:2018pqq}, so applying it to the 2d formulation of the 4d theory is straightforward now. This is the path taken in the work \cite{Oh:2019bgz}. The work \cite{Jeong:2019pzg} took a more brute force approach, explicitly Omega-deforming the HT-twisted theory in the four-dimensional terms, in the spirit of the original work \cite{Nekrasov:2002qd}.

Recall that we consider a 4d theory on $\Sigma \times C$, where $\Sigma=\C$ is the holomorphic plane, and $C$ is the $S^1$-invariant cigar supporting the B-twist. According to our earlier arguments, the Omega-deformation along $C$ can be undone in the flat region by the field redefinition. In the curved region, this leaves a non-trivial deformation coming from the superpotential \eqref{W4d2d}. In the 3d limit, $C$ is replaced by the half-line $\R_+$ ending at the $(0,4)$ boundary deformed by the boundary term $\sim \int_\Sigma \overline{W}$, as was explained earlier. The superpotential of the 2d B-model formulation \cite{Oh:2019bgz} was specified in \eqref{W4d2d}:
\begin{equation}
\cW = \int_{\Sigma} \cY \bar\partial_A \cX,
\end{equation}
where $\bar\partial_A = \dd \bar{z} (\partial_{\bar z} + A_{\bar z})$, and $\cY= q_{z}$, $\cX=\tilde{q}$ (in the twisted notations of \cite{Jeong:2019pzg}) are complex scalars of the hypermultiplet. Upon the $SU(2)_R\times SU(2)_F$-twist \cite{Jeong:2019pzg} along $\Sigma$, they can be seen as a $(1,0)$-form and a scalar along $\Sigma$. Twisting by the flavor symmetry $SU(2)_F$ is optional (and vacuous for $\Sigma=\C$), and has one advantage that $\tilde{q}, q_z$ become forms rather than spinors on $\Sigma$ \cite{Jeong:2019pzg}  The boundary term $\propto \overline{\cW}\big|_{0\in C}$ can be explicitly written as:
\begin{equation}
-2\pi i \varepsilon R^2 \int_\Sigma q_{\bar z} \dd\bar{z}\, \partial_A \tilde{q}^\dagger.
\end{equation}
In the KK reduction to 3d, one power of $R$ is absorbed into the rescaling of scalars $q$ to keep the kinetic term canonical. Thus the boundary term in 3d is:
\begin{equation}
\label{3dBdryTrm}
\tilde\varepsilon \int_\Sigma q_{\bar z} \dd\bar{z}\, \partial_A \tilde{q}^\dagger,
\end{equation}
where
\begin{align}
\tilde\varepsilon = -2\pi i \varepsilon R.
\end{align}
Finally, let us determine the supercharge in 3d. As explained around \eqref{QBdefomred}, it no longer corresponds to the 3d HT twist. The Omega-deformation along $C$ deforms the 3d HT twist to the fully topological A-twist, which is the natural deformation (since here we use $U(1)_R$ for the HT twist in 3d). Let us see it explicitly. The 3d $\cN=4$ supercharges are denoted by
\begin{equation}
{\rm Q}_{\alpha,a\dot{a}},
\end{equation}
where $\alpha=1,2$ now corresponds to the chirality along $\Sigma$, $a=1,2$ is the $SU(2)_H\equiv SU(2)_R$ index, and $\dot{a}=1,2$ is the $SU(2)_C$ R-symmetry index. There are two fully topological twists in 3d $\cN=4$ that are known under various names in the literature. One twist was originally introduced by Blau and Thompson and also by Rozansky with Witten, often going under the name of Rozansky-Witten (RW) twist \cite{Blau:1996bx,Rozansky:1996bq}, while its mirror (call it mRW) had no special name reserved. Nowadays they are often called the 3d B and A twists \cite{Kapustin:2010ag,Creutzig:2021ext,Ferrari:2023fez}, respectively, while the alternative names C and H twists are used in some literature \cite{Gaiotto:2016wcv,Coman:2023xcq}. The correspondence between them is:
\begin{align}
\text{RW twist} &= \text{B-twist} =  \text{C-twist}=\text{twist by } SU(2)_C,\cr
\text{mRW twist} &= \text{A-twist} =  \text{H-twist}=\text{twist by } SU(2)_H.
\end{align}
Each of them is fully topological, supporting a doublet of scalar supercharges given by:
\begin{align}
{\rm Q}^B_a &= \sum_{\dot{a}=1,2}{\rm Q}_{\dot{a},a \dot{a}},\cr
{\rm Q}^A_{\dot{a}} &= \sum_{a=1,2}{\rm Q}_{a,a \dot{a}}.
\end{align}

Now our 3d theory (originating from 4d on $\Sigma\times C$) lives on $\Sigma\times\R$, and $\Sigma$ supports the twist by $U(1)_H\subset SU(2)_H$. Thus there are four supercharges that are scalars along $\Sigma$:
\begin{equation}
\label{3dN4notation}
{\rm Q}_{1,1\dot{a}},\quad {\rm Q}_{2,2\dot{a}},\quad \text{where } \dot{a}=1,2.
\end{equation}
They are precisely the supercharges $Q_\pm, \overline{Q}_\pm$ of the $\cN=(2,2)$ SUSY on $C$, after $C$ has been degenerated to a line in the 3d limit. Recall that $U(1)_H=U(1)_R$ is the vector R-symmetry and $U(1)_C = U(1)_r$ is the axial R-symmetry in the $\cN=(2,2)$ algebra. Let us perform a little SUSY accounting, identifying the supercharges \eqref{3dN4notation} with $(Q_\pm, \overline{Q}_\pm)$, and indicating their R-charges:
\begin{center}
	\begin{tabular}{||c c c c||} 
		\hline
		2d $\cN=(2,2)$ & 3d $\cN=4$ & $U(1)_H$ charge & $U(1)_C$ charge \\ [0.5ex] 
		\hline\hline
		$Q_+$ & ${\rm Q}_{1,1\dot{1}}$ & 1 & 1 \\ 
		\hline
		$Q_-$ & ${\rm Q}_{1,1\dot{2}}$ & 1 & -1 \\
		\hline
		$\overline{Q}_+$ & ${\rm Q}_{2,2\dot{2}}$ & -1 & -1 \\
		\hline
		$\overline{Q}_-$ & ${\rm Q}_{2,2\dot{1}}$ & -1 & 1 \\ [1ex] 
		\hline
	\end{tabular}
\end{center}
Arbitrary linear combination of ${\rm Q}_{2,2\dot{1}}$ and ${\rm Q}_{2,2\dot{2}}$ can be chosen to define the 3d HT twist. In particular, $\cQ_B = \overline{Q}_+ + \overline{Q}_- = {\rm Q}_{2,2\dot{1}}+{\rm Q}_{2,2\dot{2}}$ is one of them. As explained around \eqref{QBdefomred}, the Omega-deformation on $C$, after the reduction to 3d, replaces this supercharge by
\begin{equation}
\label{3dTwist_following4d}
\overline{Q}_+ + \overline{Q}_- + \varepsilon(Q_+ + Q_-) = {\rm Q}_{2,2\dot{1}}+\varepsilon{\rm Q}_{1,1\dot{1}}+{\rm Q}_{2,2\dot{2}} + \varepsilon{\rm Q}_{1,1\dot{2}}.
\end{equation}
At $\varepsilon=1$, this is precisely ${\rm Q}^A_{\dot 1} + {\rm Q}^A_{\dot 2}$, one of the topological supercharges corresponding to the 3d A-twist. It is, actually, perfectly fine to keep $\varepsilon$ generic and not set it to $1$. Indeed, the $U(1)_H$ R-symmetry rotations multiply $\varepsilon$ by a phase, but one can even complexify $U(1)_H$ to $\C^\times_H$ and use it to rescale the absolute value of $\varepsilon$.

To be very clear, let us explicitly state the conclusion so far. We reduce a 4d $\cN=2$ superconformal gauge theory on the cigar $C$ supporting the $U(1)_r$-twist and the Omega-deformation, by making the cigar very thin. This flows to the 3d $\cN=4$ gauge theory with the same gauge group and matter content, now on the half-space subject to some boundary conditions. The boundary conditions are constructed by first imposing the $\cN=(0,4)$ Neumann boundary conditions on both the gauge and matter multiplets, and then deforming them by the boundary term \eqref{3dBdryTrm}. Furthermore, while without the Omega-deformation, the 4d HT twist reduces to the 3d HT twist (see Section \ref{sec:twisted_no_def}), at $\varepsilon\neq 0$ the 3d HT twist is deformed to the fully topological 3d A-twist. It is important not to confuse the twists: We perform the 2d B-twist along $C$, but end up with the 3d A-twist after the circle reduction. The following scheme may be helpful:
\begin{equation}
\label{scheme-Omega}
\boxed{\text{HT$_\varepsilon$-twist on } \Sigma \times C_\varepsilon}\quad  \vertarrowbox{\cong}{\boxed{\begin{matrix}\text{3d $\cN=4$, A-twisted on } \Sigma\times\R_+,\\ \text{with the boundary conditions }H_\varepsilon\end{matrix}}}   \quad \boxed{\text{B$_\varepsilon$-twist along } C_\varepsilon}
\end{equation}
Here the subscript $\varepsilon$ on HT and B is used to indicate the Omega-deformed version of the twist, and $C_\varepsilon$ means that the Omega-deformation is along $C$. Also, $H_\varepsilon$ is the deformed version of the boundary conditions $H$. We now proceed to study them.

\section{Deformed $\cN=(0,4)$ boundary conditions}\label{sec:DeformedBC}

In fact, the $(0,4)$ boundary conditions deformed by boundary terms have appeared in \cite{Costello:2018fnz}, and are at the heart of their construction. Here we obtain precisely the same by reduction from four dimensions. In other words, the Omega-deformation of the parent 4d HT-twisted theory reduces to the Costello-Gaiotto deformation of the $(0,4)$ boundary conditions in 3d. This is the deformation of holomorphic boundary conditions that makes them compatible with the 3d topological twist. Let us ponder this issue in more detail. We can look at this phenomenon from multiple angles.

\subsection{Review of the boundary deformation}
First, let us briefly review the deformation \cite{Costello:2018fnz}. We start with some $(0,4)$ boundary conditions $H$ in a 3d $\cN=4$ theory. The preserved $(0,4)$ supercharges are denoted as  $(\q_+, \bar{\q}_+, \q_+', \bar{\q}_+')$. At the same time, a $(2,2)$ boundary condition would preserve $(\q_+, \bar{\q}_+, \q_-, \bar{\q}_-)$. Notice how these share a pair of supercharges $(\q_+, \bar{\q}_+)$, which generate the $(0,2)$ SUSY algebra that is the intersection of the $(0,4)$ and $(2,2)$ superalgebras.

A feature of the $(2,2)$ boundary conditions is that they are compatible with the fully topological 3d $\cN=4$ twists. In particular, the 3d A-twist in the bulk corresponds to the 2d A-twist at the boundary, and same for the B-twist. This means that among the boundary-preserved supercharges, we find the bulk A and B-twist supercharges, written simply as:
\begin{equation}
{\rm Q}_A^{\rm 3d} = \bar{\q}_+ + \varepsilon\q_-,\qquad {\rm Q}_B^{\rm 3d} = \bar{\q}_+ + \varepsilon\bar{\q}_-.
\end{equation}
We allowed here for the (previously explained) possibility to have a relative coefficient $\varepsilon$ between the two supercharges, as long as $\varepsilon\neq 0, \infty$. Because the topological twist in the bulk leads to the topological twist along the boundary, we immediately see that the $(2,2)$ boundary conditions lead to topological boundary conditions in the twisted theory. They do not couple to the complex structure and cannot support any vertex algebras.

Now what about the $(0,4)$ boundary conditions? They are expected to be holomorphic. In particular, they are compatible with the HT twist, where $\bar{\q}_+$ can be taken as the HT supercharge. The $(0,4)$ boundary conditions are obviously not compatible with the topological twists in the bulk, which is bad news at first sight. What \cite{Costello:2018fnz} have noticed is that one can deform them to both be compatible with the topological twist and remain holomorphic. This is analogous to (and precedes) deforming the HT twist in the bulk to the fully topological twist \cite{Garner:2022vds}.

Holomorphic boundary conditions compatible with the topological A-twist preserve $(\rQ_A^{\rm 3d}, \q_+)$, and holomorphic boundary conditions compatible with the topological B-twist preserve $(\rQ_B^{\rm 3d}, \q_+)$. Both are special $(0,2)$ subalgebras corresponding to the $\varepsilon$-rotated embeddings of the $(0,2)$ algebra into the $(2,2)$ algebra. Both types of boundary conditions are defined, according to \cite{Costello:2018fnz}, as deformations of the $(0,4)$ boundary conditions. Let us work with the A-type supercharge for concreteness. This is the only one we need here, but the B case works analogously anyways. To define the deformation, we need to find an operator $\cO$ such that along the $(0,4)$ boundary:
\begin{align}
\label{DefDef}
\bar{\q}_+ \cO\big| = {\bS}_{-, \perp}\big| + (\text{derivative along boundary}) \neq 0,
\end{align}
where a possible derivative term drops after the boundary integration:
\begin{equation}
\bar{\q}_+ \int_\Sigma\dd^2 x\, \cO\big| = \int_\Sigma \dd^2 x\,{\bS}_{-, \perp}\big|.
\end{equation}
Here $\bS_{-,\perp}$ is the normal component of the SUSY current $\bS_{-,\mu}$ for the supercharge $\q_-$. That it does not vanish along the $(0,4)$ boundary corresponds to $\q_-$ being broken by such a boundary. At the same time, we would like to impose that
\begin{equation}
\label{ChirDef}
\q_+ \cO = \q_-\cO =0.
\end{equation}
With such conditions obeyed, the boundary insertion of
\begin{equation}
\label{BndryDefO}
e^{\varepsilon \int_{\Sigma} \dd^2 x\, \cO}
\end{equation}
in the path integral (or inside all the correlators) provides the necessary deformation. The supercharge $\q_+$ remains conserved, since it is preserved by the undeformed boundary $H$, and $\q_+ \cO=0$ ensures that the deformation does not break it either. The deformation breaks $\bar{\q}_+$ SUSY by the amount $\varepsilon \int_\Sigma \dd^2 x\, \bar{\q}_+ \cO$. On the other hand, the $\q_-$ supercharge is broken by $H$, and the rate of breaking is given by the odd flux $-\int_\Sigma \dd^2x\, \bS_{-,\perp}$ through the boundary. This breaking is not affected by the boundary deformation, since $\q_-\cO=0$. Thus we see that the non-conservations of $\bar{\q}_+$ and $\q_-$ (i.e., the odd flux describing the rate at which these supercharges leak through the boundary) are proportional to each other, and the new conserved supercharge is $\rQ_A^{\rm 3d} = \bar{\q}_+ + \varepsilon \q_-$. Therefore, the deformed boundary preserves $(\rQ_A^{\rm 3d}, \q_+)$, which is a special $(0,2)$ subalgebra, since
\begin{equation}
\{\rQ_A^{\rm 3d}, \q_+\} = 2\bar\partial.
\end{equation}
In other words, we obtain holomorphic boundary conditions compatible with the topological A-twist in the bulk. The B-twist version works verbatim, with the supercharge $\rQ_A^{\rm 3d}$ replaced by $\rQ_B^{\rm 3d}=\bar{\q}_+ + \varepsilon\bar{\q}_-$.

\subsection{The 2d B-model viewpoint on the deformation}
Let us go back to the 2d B-model on $C$, which was used in Section \ref{sec:BtwistOmega} to formulate the Omega deformation. The B-model supercharge,
\begin{equation}
\cQ_B = \overline{Q}_+ + \overline{Q}_-,
\end{equation}
was Omega-deformed,
\begin{equation}
\cQ_B + \varepsilon \iota_V G,
\end{equation}
and in the flat region of cigar it was equal to \eqref{QBdefomred}:
\begin{equation}
\overline{Q}_+ + \overline{Q}_- + \varepsilon(Q_+ + Q_-).
\end{equation}
Let us look at the supersymmetry current for $Q_+ + Q_-$. Its radial component on $C$ becomes normal component at the effective $(0,4)$ boundary. For example, in a 2d $\cN=(2,2)$ LG model, this normal component is:
\begin{equation}
\left(S_{+,\perp} + S_{-,\perp}\right)\big| = \frac{i}{2}(\psi_+^i\bar{\partial}\bar{x}^i + \psi_-^i\partial \bar{x}^i)\big| + i (\bar\psi_-^i + \bar\psi_+^i)\partial_i\overline{\cW}\big| = i (\bar\psi_-^i + \bar\psi_+^i)\partial_i\overline{\cW}\big|,
\end{equation}
where the first term vanishes via the effective boundary conditions, in the 1d limit of $C$ degenerating to $\R_+$. Recognizing the last term as $\cQ_B \overline{\cW}$, we see that
\begin{equation}
\cQ_B \overline{\cW}\big| = \left(S_{+,\perp} + S_{-,\perp}\right)\big|.
\end{equation}
This is precisely the relation \eqref{DefDef} seen through the 2d B-model eyes, implying that the boundary term $\varepsilon \overline{\cW}$ deforms the $\cQ_B$ supercharge to
\begin{equation}
\cQ_B + \varepsilon (Q_+ + Q_-).
\end{equation}
This manifestly generalizes to the gauged 2d B-model, with the boundary term \eqref{3dBdryTrm} and the same conclusion.

This reasoning shows that $\overline{\cW}$, --- the complex conjugate of the 2d superpotential, --- is the canonical choice for the boundary deformation $\int\dd^2x\, \cO$ in \eqref{BndryDefO}, at least when the 2d formulation in terms of the B-model on $C$ is available. It is only useful for Lagrangian theories, so it is good to have other characterizations of $\cO$ that apply to more general non-Lagrangian examples.

\subsection{The 3d stress-energy multiplet viewpoint}
Let us now look at the boundary deformation from the purely three-dimensional point of view. The key relation \eqref{DefDef} suggests that $\cO$ must be part of the same SUSY multiplet as the supersymmetry current. This is the supercurrent multiplet, also called the stress-energy tensor multiplet, since the stress-energy tensor resides in it as well. The general 3d $\cN=4$ stress-energy multiplet does not seem readily available in the literature, though perhaps it can be extracted with some work. For simplicity, let us assume that the 3d theory is an SCFT. This is a good assumption, since most of our 3d theories flow to some SCFT in the IR. Unlike in 4d, it does not provide that significant of a limitation in 3d.

The structure of the 3d $\cN=4$ superconformal stress-tensor multiplet can be read off from the literature with significantly less effort, see for example \cite{Bergshoeff:2010ui}. This multiplet is unique. It contains components $(a, \rho_i, \sigma, J_\mu^{ij}, S_\mu^i, T_{\mu\nu})$. Here $a$ is a scalar of zero R-charge and conformal dimension $1$; $\rho_i$ is a spinor of R-charge $(\frac12, \frac12)$ under $SU(2)_H\times SU(2)_C$ and conformal dimension $\frac23$; $\sigma$ is an R-neutral scalar of conformal dimension $2$; $J_\mu^{ij}$ is the R-symmetry current of conformal dimension $2$; $S^i_\mu$ is the supercurrent of R-charge $(\frac12, \frac12)$ and conformal dimensions $\frac32$; $T_{\mu\nu}$ is the stress-energy tensor. Here we use the $SO(4)$ notations, so $i,j=1,..,4$. The SUSY variations are:
\begin{align}
\label{stress-scft}
\delta T_{\mu\nu} &= -\frac14 \bar\epsilon^i \gamma_{(\mu}{}^\rho \partial_\rho S_{\nu)}{}^i,\cr\delta S_{\mu}^i &=\gamma^\nu\epsilon^i T_{\mu\nu} + \varepsilon_{\mu\nu\rho}\gamma^\sigma \gamma^\nu \epsilon_j \partial^\rho J_\sigma^{ij},\cr
\delta J_\mu^{ij} &= \frac{i}{2}\bar\epsilon^{[i}S_\mu^{j]} + \bar\epsilon_k \varepsilon^{ijkl}\gamma_{\mu\nu}\partial^\nu \rho_l,\cr
\delta\sigma &= \bar\epsilon_i \gamma^\mu\partial_\mu \rho_i,\cr
\delta\rho_i &= \frac18 \varepsilon_{ijkl}\gamma^\mu \epsilon^l J^{jk}_\mu +\frac14 \sigma\epsilon_i -\frac14 \gamma^\mu\partial_\mu a \epsilon_i,\cr
\delta a &= -\bar\epsilon^i\rho_i.
\end{align}
Superconformal invariance implies that, in addition to the conservation laws, $T_{\mu\nu}$ is traceless, and $S_\mu^i$ is $\gamma$-traceless:
\begin{equation}
T_\mu{}^\mu = 0,\quad \gamma^\mu S_\mu^i=0.
\end{equation}
One can check explicitly that for the Poincare SUSY transformations, the variations in \eqref{stress-scft} close according to the super-Poincare algebra:
\begin{equation}
\{\delta_{\epsilon_2}, \delta_{\epsilon_1}\} = \frac12 \bar{\epsilon}_1^i \gamma^\mu \epsilon_2^i \partial_\mu,
\end{equation}
but only when the following constraints hold:
\begin{align}
\partial^\mu T_{\mu\nu}&=0,\quad \partial^\mu S_\mu^i=0,\quad \partial^\mu J_\mu^{ij}=0,\cr
T_\mu{}^\mu&=0,\quad \gamma^\mu S_\mu^i=0.
\end{align}
In the first line, we see the usual conservation laws, while the second line contains conditions implying the superconformal symmetry. Thus the stress-tensor multiplet \eqref{stress-scft} indeed only applies to superconformal theories. One can further check that for the $\epsilon_i$'s representing the superconformal transformations, \eqref{stress-scft} also closes, now according to the superconformal algebra.

It may be also useful to rewrite the multiplet in the $SU(2)_H\times SU(2)_C$ indices, rather than $SO(4)_R$. For that, introduce the intertwining operators:
\begin{align}
\sigma_i^{a\dot{a}} &= (\sigma_0=1, \sigma_{j=1,2,3}=i\tau^j),\cr
\tilde\sigma^i_{\dot{a}a} &= (\tilde\sigma^0=1, \tilde\sigma^{j=1,2,3}=-i\tau^j),
\end{align}
where $\tau^j$ are Pauli matrices. Here $x^i \sigma^i$ represents a quarternion, and the $SU(2)_H\times SU(2)_C$ acts on it via left and right multiplications by the unit quarternions:
\begin{equation}
(\xi_a, \zeta_a): (x^i \sigma^i) \mapsto e^{i \xi_a \tau^a} (x^i \sigma^i) e^{-i \zeta_a \tau^a }.
\end{equation}
Let us identify $SU(2)_H$ with the left action and $SU(2)_C$ -- with the right. One can show that the left and right generators are built from $J^{ij}$ via:
\begin{align}
(J^H)^a{}_b=(J^L)^a{}_b &= -\frac{i}{2} J^{ij} \sigma_i^{a\dot{a}} \tilde{\sigma}^j_{\dot{a}b},\cr
(J^C)_{\dot a}{}^{\dot b}=(J^R)_{\dot a}{}^{\dot b} &= -\frac{i}{2} J^{ij} \tilde{\sigma}^i_{\dot{a}a} {\sigma}_j^{a\dot{b}}.
\end{align}
We note that
\begin{equation}
\sigma_{ij} = \sigma_{[i}\tilde{\sigma}_{j]}
\end{equation}
is a self-dual tensor. We also raise and lower indices according to $v^a=\varepsilon^{ab}v_b$. In addition, one can check that:
\begin{equation}
\tilde{\sigma}^i_{\dot{a}a} = \varepsilon_{ab}\varepsilon_{\dot{a}\dot{b}} \sigma_i^{b\dot{b}}.
\end{equation}
Using these rules, we find:
\begin{equation}
\delta (J^H_{\mu})^{ab} = \frac14 \bar{\epsilon}^{(a|\dot{a}}S_{\mu}{}^{b)}{}_{\dot a} - i\bar{\epsilon}^{(a|\dot{a}}\gamma_{\mu\nu}\partial^\nu \rho^{b)}{}_{\dot a}.
\end{equation}

The 3d HT supercharge $\bar{\q}_+$ in our case follows from the 4d HT twist, which according to \eqref{3dTwist_following4d} is $\rQ_{2,2\dot{2}} + \rQ_{2,2\dot{1}}$ in the 3d notations. The $SU(2)_C$ rotations make it into a general linear combination $c_{\dot a} \rQ_2^{2\dot{a}}$, which still defines a valid HT supercharge. For convenience of computations, let us temporarily take $\rQ_{2,2\dot{2}}$ as the HT supercharge. Then $\q_+$ is $\rQ_{2,1\dot{1}}$ and $\q_-$ is $\rQ_{1,1\dot{2}}$. We want ot realize the deformation $\bar{\q}_+ + \varepsilon \q_-$, so the relevant supercurrent component entering the condition \eqref{DefDef} is $S_{1\perp,1\dot{2}}$. We then compute:
\begin{equation}
\label{QplusJ}
Q_{2,2\dot{2}} (J^H_z)^{22} = \frac14 S_{2z,1\dot{2}} + i \partial_z \rho_{2,1\dot{2}} = -\frac18 S_{1\perp,1\dot{2}} + i \partial_z \rho_{2,1\dot{2}},
\end{equation}
where the last equality utilizes the gamma-tracelessness of $S$, which can be written as $\gamma^\mu S_{\mu,a\dot{a}}=0$. Upon integration over $\Sigma$, the derivative term drops out and we obtain:
\begin{equation}
Q_{2,2\dot{2}}\int_\Sigma \dd^2 z\, (J^H_z)^{22} = -\frac18 \int_\Sigma \dd^2z\, S_{1\perp,1\dot{2}}.
\end{equation}
It is also easy to check that:
\begin{equation}
\label{QQJ}
Q_{2,1\dot{1}} (J^H_z)^{22} = Q_{1,1\dot{2}} (J^H_z)^{22}=0.
\end{equation}
Thus $(J^H_z)^{22}$ satisfies all the requirements for the boundary deformation $\cO$. Furthermore, the explicit form of the boundary deformation in gauge theories, which we found in \eqref{3dBdryTrm},\footnote{Here we use the notation $(q^a, \tq^a)$ for the hypermultiplet scalars, which was not used in previous sections. Here $a=1,2$, $q^1$ and $\tq^1$ are the bottom components of the chiral multiplets, and $(q^a)^* = \tq_a$. In the twisted notations used around \eqref{3dBdryTrm}, $q_z$ is $q^1$, $\tq$ is $\tq^1$, $q_{\bar z}$ is $-\tq^2$, and $\tq^\dagger$ is $q^2$.}
\begin{equation}
\cO = \tq^2 \partial_A q^2,
\end{equation}
precisely matches the bosonic part of component $(J^H_z)^{22}$ of the $SU(2)_H$ R-symmetry current. The fermionic terms in $(J^H_z)^{22}$ vanish at the $(0,4)$ boundary $H$. Thus we verify that the boundary deformation relevant for the H (or 3d A) twist is
\begin{equation}
e^{\varepsilon \int (J_z^H)^{22} \dd^2 x}.
\end{equation}
The analogous deformation relevant for the C (or 3d B) twist is given by the component $(J_z^C)^{\dot{2}\dot{2}}$ of the $SU(2)_C$ current. Such an explicit description of $\cO$ is also valid for non-Lagrangian SCFTs.

\subsection{The 4d stress-energy multiplet viewpoint}
Above we have explicitly identified the boundary deformation as the component $(J_z^{\rm H})^{22}$ of the R-symmetry current, but the argument was limited to 3d SCFTs only. In this paper, our starting point is a 4d SCFT, though. After the circle reduction, it does not immediately become the 3d SCFT, -- it still has to flow to the IR. We can slightly improve on this situation by instead relying on the superconformal 4d $\cN=2$ stress-tensor multiplet in our argument. It is extracted from the Sohnius multiplet \cite{Sohnius:1978pk} when the submultiplet of anomalies vanishes. In particular, the SUSY transformation of the $SU(2)_R$ current involves the supersymmetry currents $S_{\mu}^a$, $\widetilde{S}_{\mu, a}$ and the additional $SU(2)_R$-doublet of dimension $\frac52$ Weyl spinors $\chi^a$, $\overline{\chi}_a$:
\begin{equation}
\delta J_{\mu}^{ab} = \epsilon^{(a} S_{\mu}^{b)} + \overline{\epsilon}^{(a} \overline{S}_{\mu}^{b)} -\frac{i}{3} \epsilon^{(a} \sigma_{\mu\nu}\partial^\nu \chi^{b)} -\frac{i}{3} \overline{\epsilon}^{(a} \overline{\sigma}_{\mu\nu}\partial^\nu \overline{\chi}^{b)}
\end{equation}
We unravel this to compute several 4d SUSY variations of $J_{z}^{22}$:
\begin{align}
\rQ^2_\pm J_{z}^{22} &= \tQ_{1,\dot{\pm}} J_z^{22}=0,\cr
\rQ^1_- J_{z}^{22} &= 
-S^2_{z-} -\frac{i}{3} \partial_z \chi^2_{-},\cr
\tQ_{2\dot{-}} J_z^{22} &= 
\widetilde{S}^2_{z\dot{-}} + \frac{i}{3} \partial_z\overline{\chi}^2_{\dot-}
\end{align}
Because $\q_+=\tQ_{1\dot{-}} + Q^2_-$ and $\q_-=\tQ_{1\dot{+}} + Q^2_+$, we immediately see that
\begin{equation}
\q_\pm J_z^{22} = 0.
\end{equation}
Thus two conditions are obeyed. As for $\bar{\q}_+=Q^1_- + \tQ_{2\dot{-}}$, we also need the gamma-tracelessness $\gamma^\mu S_\mu^a = \gamma^\mu \widetilde{S}_{\mu,a}=0$, which allows to express:
\begin{equation}
S^2_{z-}=-S^2_{\bar{w}+},\qquad \widetilde{S}^2_{z\dot{-}}=-\widetilde{S}^2_{w\dot{+}},
\end{equation}
where $z$ is the holomorphic coordinate on $\Sigma$, and $w$ -- on $C$. This gives:
\begin{equation}
\label{supercurr4d_res}
\bar{\q}_+ J_z^{22} = S^2_{\bar{w}+} + \widetilde{S}_{w\dot{+},1} +\frac{i}{3} \partial_z (\chi^2_{-}+\overline{\chi}^2_{\dot-}).
\end{equation}
Again, the derivative term drops after the integration over $\Sigma$. The condition \eqref{DefDef} is almost obeyed, except the supercurrent has some unwanted terms. Of course \eqref{DefDef} is written in 3d, while \eqref{supercurr4d_res} is in 4d, so we expect the unwanted terms to vanish in the 3d limit. In the cylindrical region of $C$, where $r$ is the longitudinal coordinate and $\varphi$ is the radial one, pick:
\begin{equation}
w=r +i\varphi.
\end{equation}
Then in such coordinates:
\begin{equation}
S^2_{\bar{w}+}=\frac12(S^2_{r+} + iS^2_{\varphi+}),\qquad \widetilde{S}_{w\dot{+},1}=\frac12(\widetilde{S}_{r\dot{+},1} - i\widetilde{S}_{\varphi\dot{+},1}).
\end{equation}
In the 3d limit, the angular components $S^2_{\varphi+}$ and $\widetilde{S}_{\varphi\dot{+},1}$ vanish, while the longitudinal components become ``$\perp$'' near the boundary, so $S^2_{r+} + \widetilde{S}_{r\dot{+},1}$ is indeed the normal component of the current for $\q_-$. Thus the condition \eqref{DefDef} is obeyed in the 3d limit.

\textbf{Remark:} We have shown that the component $J_z^{22}$ of the $SU(2)_R=SU(2)_H$ R-symmetry current provides the necessary boundary deformation. Notice that the lowest component $J_z^{11}$ of the same current is, famously, the Schur operator that descends to the stress-energy tensor in the 2d chiral algebra \cite{Beem:2013sza}. The twisted translations of \cite{Beem:2013sza} mix $J_z^{11}$ with $J_z^{12}=J_z^{21}$ and $J_z^{22}$. In the current paper, the twisted translations are no longer necessary, their job is done by the Omega-deformation. Nevertheless, the component $J_z^{22}$ still shows up, now as the boundary deformation term in the 3d description. This seems like a curious observation, but we are not aware if it has any interesting consequences.

\section{3d TQFT and rank zero SCFT}\label{sec:tqft}
\subsection{TQFT and conformal blocks}\label{sec:confblck}
The boundary VOA is compatible with the H-twist (topological 3d A-twist) in the bulk by construction. We apply the topological A-twist to the 3d theory. This gives a cohomological 3d TQFT in the bulk coupled to the holomorphic boundary conditions \cite{Costello:2018fnz}. We can pick some state $\Psi\in \cH[\Sigma]$ from the TQFT Hilbert space on $\Sigma$ and evaluate it against the boundary. One can also decorate the boundary with insertions of the boundary vertex operators, and compute correlation functions $\langle \cO_1 \cO_2\dots \cO_n \rangle_\Psi$ ``in the state'' $\Psi$, see Figure \ref{fig:tqft-block}:\newpage
\begin{figure}[h]
	\centering
	\includegraphics[scale=0.5]{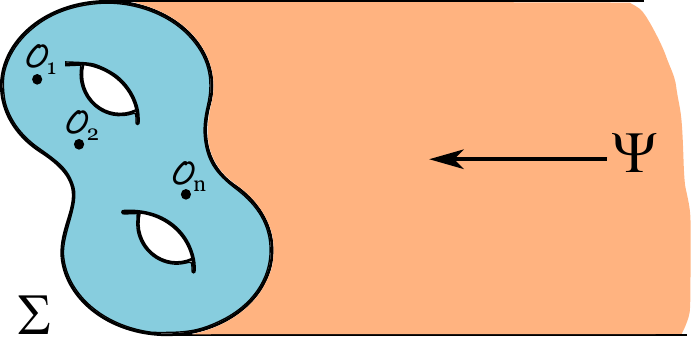}
	\caption{State $\Psi$ in the TQFT provides conformal block for the boundary VOA.\label{fig:tqft-block}}
\end{figure}
Such correlators are obviously compatible with the VOA algebraic structure, as well as conformal invariance. The state $\Psi$ is therefore interpreted as a conformal block for the boundary VOA \cite{Gaiotto:2016wcv}, and $\cH[\Sigma]$ -- as the subspace in the space of all such blocks. In many cases, $\cH[\Sigma]$ in TQFT is known to coincide with the space of conformal blocks: This is quite familiar for the Chern-Simons theories \cite{Witten:1988hf,Elitzur:1989nr}, or more generally, 3d TQFT corresponding to RCFT \cite{Fjelstad:2005ua,Runkel:2005qw,Kapustin:2010if}. As for the non-unitary (cohomological) TQFTs that are constructed via the 3d topological twist, this picture is less mature at the moment. The boundary VOA of \cite{Costello:2018fnz} provides an important step towards the general understanding. The module categories of such VOAs are related to the categories of lines in the TQFT, allowing to recover information on the moduli spaces of vacua \cite{Costello:2018swh}. The categories may be non-semisimple (see e.g., \cite{Creutzig:2021ext,Gukov:2020lqm,Costantino:2021yfd,Cheng:2018vpl,CGP}), and the VOAs do not have to be rational or $C_2$-cofinite. See talks \cite{pirsa_PIRSA:20020075,pirsa_PIRSA:20020084} where some ideas along these lines are formulated (and \cite{Witten:2010zr,Witten:2010cx} for the background).

The 3d TQFT comes with the mapping class group ${\rm MCG}(\Sigma)$ acting on $\cH[\Sigma]$, something that requires proof when looked purely from the VOA side. Therefore, the subspace of conformal blocks $\cH[\Sigma]$ must be closed under such actions. One particularly important case is $\Sigma = \bT^2$, where we compute VOA characters decorated by the vertex operator insertions. It was already proven in \cite{Zhu} that the vector space of torus conformal blocks is finite-dimensional for the $C_2$-cofinite VOAs. It was also shown that for a rational VOA, it is spanned by the irreducible modules (i.e., in a given conformal block, torus correlators are defined via traces over the corresponding module). Modular invariance in the rational case was also proven there (for the quasi-lisse case see \cite{Arakawa:2016hkg}, also see the collection of references in \cite{Dedushenko:2019mzv}).

So simply because we have a TQFT, the space of torus blocks $\cH[\bT^2]$ is closed under the modular PSL$(2,\Z)$ action. Among such blocks, there is always one corresponding to the vacuum module, whose torus correlators are defined as the vacuum character with insertions. The state $|0\rangle\in \cH[\bT^2]$ corresponding to the vacuum block is associated by our TQFT to the empty solid torus, as on Figure \ref{fig:vac-block}:\newpage
\begin{figure}[h]
	\centering
	\includegraphics[scale=0.5]{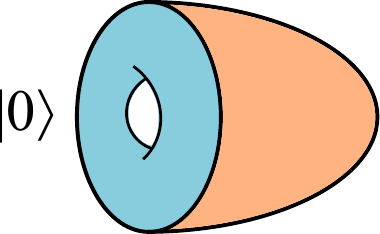}
	\caption{State $|0\rangle$ corresponding to the vacuum block is generated by the empty solid torus.\label{fig:vac-block}}
\end{figure}
It is proven using the standard arguments. First, the boundary version of the state-operator map implies that the vacuum module $V$ of the boundary VOA is the space of states on the disk $\cH[D^2]$, with our holomorphic boundary conditions $H_\varepsilon$, see Figure \ref{fig:radial}:
\begin{figure}[h]
	\centering
	\includegraphics[scale=0.5]{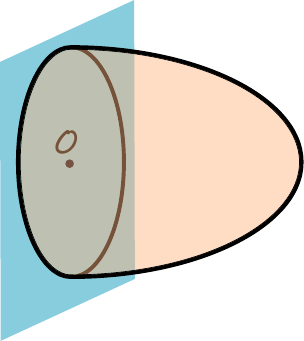}
	\caption{Vertex operator $\cO \in V$ is inserted at the origin of the boundary $H_\varepsilon$. It creates a state $|\cO\rangle \in \cH[D^2]$ in the Hilbert space associated to the disk ending on the boundary $H_\varepsilon$.\label{fig:radial}}
\end{figure}\\
Thus the vacuum character is the trace over $\cH[D^2]$, and so is given by the TQFT partition function on the empty solid torus with the $H_\varepsilon$ boundary conditions. Decorating vacuum character by the boundary insertions is now straightforward. We start with some state in the vacuum module, $a \in \cH[D^2]$, evolve it with $q^{L_0}$ (this evolution is only nontrivial along the boundary, since the bulk is topological), adding some boundary insertions along the way, and cap it off with $a$ again, to compute the trace. The Figure \ref{fig:decor-tor} is self-explanatory:
\begin{figure}[h]
	\centering
	\includegraphics[scale=0.5]{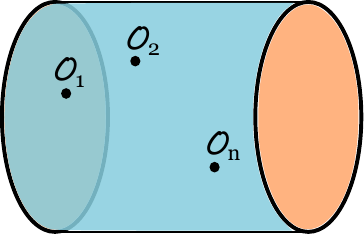}
	\caption{Boundary insertions $\cO_1, \dots, \cO_n$ in the configuration computing the vacuum character. Here the left and right orange disks $D^2$ are glued together, corresponding to taking the trace over $\cH[D^2]$. This produces a solid torus with the boundary insertions.\label{fig:decor-tor}}
\end{figure}\\
So torus correlators $\langle \cO_1\dots \cO_n\rangle_0$ for the vacuum module are computed by the empty solid torus with $\cO_1,\dots, \cO_n$ inserted along the boundary. This proves that the state $|0\rangle$ created by the empty solid torus is the vacuum conformal block. Under the modular action, $|0\rangle$ is mapped to other states in $\cH[\bT^2]$. The topological state-operator map identifies $\cH[\bT^2]$ with the space of line operators $\cL$ at the center of the solid torus, see Figure \ref{fig:line-block}:
\begin{figure}[h]
	\centering
	\includegraphics[scale=1]{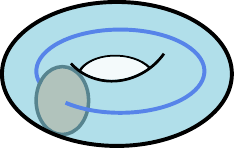}
	\caption{Line operator $\cL$ at the center of the solid torus creates a state $|\cL\rangle\in \cH[\bT^2]$.\label{fig:line-block}}
\end{figure}\\
When the line $\cL$ ends at the holomorphic boundary $H_\varepsilon$, it creates the module $M_\cL$ for the boundary VOA $V$. Using the same argument as for the vacuum block, we immediately see that correlators in the torus conformal block $|\cL\rangle$ are simply computed by the character of $M_\cL$, decorated by the vertex operator insertions. In the standard TQFT/RCFT dictionary, there exists a finite set of simple lines/anyons $\cL$, such that $M_\cL$ are simple modules for the VOA. For the non-unitary non-semisimple TQFTs we may encounter, the dictionary is expected to be less trivial. See for instance an example of $V_{-\frac{-4n}{2n+1}}\left(\mathfrak{su}(2)\right)$ in \cite{Beem:2017ooy}, where the non-vacuum characters one encounters are quite peculiar. For such admissible level affine $\mathfrak{su}(2)$, there is an interesting theory of modular invariance developed in \cite{Creutzig:2012sd,Creutzig:2013yca}, which significantly enlarges the module category. It would be useful to understand it from the 3d TQFT point of view, since such VOAs appear at the boundaries of 3d topological theories descending from the $(A_1, D_{2n+1})$ Argyres-Douglas theories, as we will see below.

Also note that since our 3d theories originate in 4d, the line operators $\cL$ are expected to come from the surface operators in 4d wrapping the topological cigar $C$, see Figure \ref{fig:surf-line}:
\begin{figure}[h]
	\centering
	\includegraphics[scale=1]{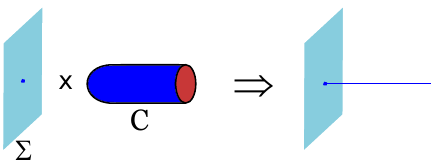}
	\caption{A surface operator wrapping $C$ reduces to the line operator terminating on the holomorphic boundary.\label{fig:surf-line}}
\end{figure}\\
Such surfaces must preserve $\cN=(2,2)$ SUSY \cite{Beem:2013sza} to give modules for the chiral algebra, and indeed this amount of SUSY is compatible with the B-twist along $C$ that we perform.

\subsection{3d TQFT from $S^1$ reduction with a $\Z_N$ twist}\label{sec:3dTQFT}
What kind of 3d TQFTs do we have to deal with? They originate from the $S^1$ reduction of 4d SCFTs, followed by the topological A-twist in 3d. When the parent theory is Lagrangian, we simply obtain the twisted version of the ``same'' Lagrangian gauge theory in 3d, with the same gauge group and matter content as in 4d. Boundary VOAs in such theories have recently received some attention in the literature \cite{Costello:2018fnz,Costello:2018swh,Creutzig:2021ext,Ballin:2022rto,Garner:2022rwe,Yoshida:2023wyt,Beem:2023dub,Ferrari:2023fez,Coman:2023xcq}.

Let us look at another class of examples, -- the non-Lagrangian theories with fractional $U(1)_r$ charges. Recall the issue discussed at the end of Section \ref{sec:HTtwist}. Namely, the $U(1)_r$ twist along $C$ creates the monodromy $e^{2\pi i r}$ around the asymptotic circle of $C$, for the fields of r-charge $r$.
In theories with fractional $U(1)_r$ charges, such as the Argyres-Douglas (AD) \cite{Argyres:1995jj,Argyres:1995xn,Bonelli:2011aa,Xie:2012hs,Xie:2013jc} ones, this leads to an interesting observation. An operator of r-charge $\frac{n}{N}$ acquires a non-trivial phase $e^{2\pi i \frac{n}{N}}$ under the total $e^{2\pi i}\in U(1)_r$ rotation. Thus $e^{2\pi i}\in U(1)_r$, though commuting with the supercharges, acts as a non-trivial global symmetry transformation. If $N$ is the lowest common denominator of all the fractional r-charges, then it is clear that $e^{2\pi i}\in U(1)_r$ generates a global $\Z_N$ symmetry. In other words, we find that $U(1)_r$ is extended by $\Z_N$ to $\widetilde{U(1)}_r$, which is the N-fold covering of $U(1)_r$, as manifested by the short exact sequence:
\begin{equation}
1 \longrightarrow \Z_N \longrightarrow \widetilde{U(1)}_r \longrightarrow U(1)_r \longrightarrow 1.
\end{equation}
Our cigar, due to the topological twist, carries precisely the holonomy $e^{2\pi i}\in U(1)_r$ along the asymptotic circle, see Figure \ref{fig:r-holo}:
\begin{figure}[h]
	\centering
	\includegraphics[scale=1]{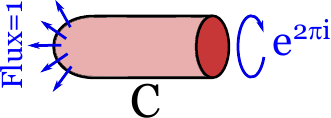}
	\caption{Topological twist creates a unit r-symmetry holonomy along the asymptotic circle of $C$, which acts non-trivially on the operators with fractional r-charges.\label{fig:r-holo}}
\end{figure}\\
Thus a 4d SCFT with fractional r-charges is reduced on the $S^1$ with special periodicity conditions. They are given by the standard supersymmetric (i.e., periodic) conditions modified by the additional discrete holonomy around this $S^1$:
\begin{equation}
\label{discr_holo}
\gamma = (1\mod N) \in \Z_N.
\end{equation}
Such discrete holonomies in AD theories were first discussed in \cite{Dedushenko:2018bpp}. They lead to interesting 3d SCFTs in the IR limit, which bear no manifest relation to the well-understood case of the ordinary periodic $S^1$ reduction of the AD theories \cite{Xie:2012hs,Buican:2015hsa,Dedushenko:2019mzv,Dedushenko:2019mnd,Closset:2020afy,Giacomelli:2020ryy,Xie:2021ewm}.

This $\Z_N$, unlike continuous flavor symmetries, acts non-trivially on the Coulomb branch $\cM$ of the 4d SCFT on $S^1\times \R^3$. In the class S theories, $\cM$ is given by the appropriate Hitchin moduli space $\cM_H$ (with ramifications) \cite{Seiberg:1996nz,Gaiotto:2010okc,Gaiotto:2009hg}. It was shown in \cite{Dedushenko:2018bpp} that under a special \emph{non-integrality condition}, the $\Z_N$-fixed locus of $\cM$ coincides with the $\widetilde{U(1)}_r$-fixed locus, and is given by a discrete collection of points. This non-integrality condition is easily stated: No generators of the 4d Coulomb branch chiral ring can have integral r-charges (or scaling dimensions, since $\Delta=r$ for the Coulomb branch operators).

Let us briefly review the main steps. First, $\cM$ is $4k$-dimensional, hyper-K{\"a}hler, with complex structures $(I,J,K)$ \cite{Seiberg:1996nz,Gaiotto:2010okc}. Say in the complex structure $I$, it is holomorphically fibered over the $2k$-dimensional (or $k$-complex-dimensional) base $\cB$:
\begin{equation}
\pi: \cM \to \cB.
\end{equation}
The fibers of $\pi$ are Lagrangian with respect to $\Omega_I=\omega_J + i\omega_K$. The $\widetilde{U(1)}_r$ acts on both $\cM$ and $\cB$, and $\pi$ is $\widetilde{U(1)}_r$-equivariant. This $\cB$ is the Coulomb branch on $\R^4$, -- the affine variety, whose coordinate ring is the 4d Coulomb branch chiral ring.

Because of the non-integrality condition, the $\Z_N$-fixed locus of $\cB$ is the origin, $\cB^{\Z_N}=\{0\}$. This implies that $\cM^{\Z_N}\subset \pi^{-1}(0)$. Being the flavor symmetry, $\Z_N$ preserves the hyper-K{\"a}hler structure, so $\cM^{\Z_N}$ is hyper-K{\"a}hler as well. In particular, it is symplectic with respect to $\Omega_I$, while at the same time it lies within the fiber $\pi^{-1}(0)$ that is Lagrangian with respect to $\Omega_I$. These two facts clash with each other, unless $\cM^{\Z_N}$ is zero-dimensional, i.e., a collection of points. Furthermore, $\cM^{\Z_N} = \cM^{\widetilde{U(1)}_r}$. The inclusion $\cM^{\widetilde{U(1)}_r}\subset\cM^{\Z_N}$ is trivial, since $\Z_N$ is a subgroup of $\widetilde{U(1)}_r$. In the other direction: If a point is fixed by $\Z_N$, it better be fixed by $\widetilde{U(1)}_r$ too, or else we would get a continuous orbit of $\Z_N$-fixed points, which contradicts the zero-dimensionality of $\cM^{\Z_N}$. So we proved that if the non-integrality condition is obeyed, $\cM^{\Z_N} = \cM^{\widetilde{U(1)}_r}$ is simply a discrete collection of fixed points.

In the 3d limit, $\cM$ degenerates to the Coulomb branch of the 3d SCFT. When we turn on the holonomy \eqref{discr_holo}, it is $\cM^{\Z_N}$ that degenerates to the Coulomb branch of the 3d SCFT. Indeed, only the operators that do not have monodromy around the $S^1$ can acquire a vev in the Coulomb branch vacuum, and they precisely parameterize $\cM^{\Z_N}$. Thus we conclude: In theories obeying the non-integrality condition, the Coulomb branch is lifted in the 3d limit by the discrete holonomy \eqref{discr_holo}. Only a discrete set of points is left behind.

What theories obey the non-integrality condition? The best bet are Argyres-Douglas theories. For the generalized AD theories \cite{Shapere:1999xr,Cecotti:2010fi,Cecotti:2011rv,Bonelli:2011aa,Xie:2012hs,Xie:2013jc,Wang:2015mra,Wang:2018gvb,Beem:2023ofp}, it is not hard to extract the spectrum of Coulomb branch operators. For example, in the standard series:
\begin{align}
(A_1, A_{2n}):  \quad \Delta &= 1+ \frac{1}{2n+3}, 1+ \frac{3}{2n+3}, \dots, 2- \frac{4}{2n+3},\quad N=2n+3 \cr
(A_1, A_{2n-1}): \quad \Delta &= 1+ \frac{2}{2n+2}, 1+ \frac{4}{2n+2}, \dots, 2-\frac{4}{2n+2},\quad N=2n+2  \cr
(A_1, D_{2n+1}): \quad \Delta &=1+\frac{1}{2n+1}, 1+\frac{3}{2n+1}, \dots, 2 - \frac{2}{2n+1},\quad N=2n+1 \cr
(A_1, D_{2n}): \quad \Delta &= 1+\frac{1}{n}, 1+\frac{2}{n}, \dots, 2-\frac{1}{n},\quad N=n.
\end{align}
We see that all the Coulomb branch generators have dimensions between $1$ and $2$, that is $1<\Delta=r <2$. Thus the non-integrality condition is obeyed, and for such theories the $\Z_N$-twisted circle reduction fully removes the Coulomb branch. That is, we end up with 3d $\cN=4$ theories without Coulomb branches.

We could look at other series of AD theories, for example \cite{Xie:2012hs}:
\begin{align}
(A_2, A_{3n}):  \quad \Delta &= 1+ \frac{1}{3n+4}, 1+ \frac{4}{3n+4}, \dots, 2- \frac{6}{3n+4},\quad N=3n+4 \cr
\Delta&= 1+ \frac{2}{3n+4}, 1+\frac{5}{3n+4}, \dots, 3-\frac{9}{3n+4},\cr
(A_2, A_{3n-1}): \quad \Delta &= 1+ \frac{1}{n+1}, 1+ \frac{2}{n+1}, \dots, 2-\frac{2}{n+1},\quad N=n+1  \cr
\Delta&=1 + \frac{1}{n+1}, 1+ \frac{2}{n+1}, \dots, 3-\frac{3}{n+1} \cr
(A_2, A_{3n-2}): \quad \Delta &=1+\frac{2}{3n+2}, 1+\frac{5}{3n+2}, \dots, 2 - \frac{6}{3n+2},\quad N=3n+2\cr
\Delta &= 1 + \frac1{3n+2}, 1+ \frac4{3n+2}, \dots, 3 - \frac{9}{3n+2}.
\end{align}
Here each of the $(A_2, A_M)$ theories has two sets of the Coulomb operators, and we write sequences of their dimensions in the two rows. Each sequence is an arithmetic progression. Clearly, the theories $(A_2, A_{3n})$ and $(A_2, A_{3n-2})$ have no integers in these sequences, and so obey the non-integrality. At the same time, the theory $(A_2, A_{3n-1})$ for $n\geq 2$ contains $\Delta=2$. Thus it does not obey the non-integrality, precisely one Coulomb branch operator violating it. In the 3d limit, therefore, this operator survives. Together with the Hitchin fiber, one finds four-real-dimensional Coulomb branch in the 3d limit.

This illustrates that the problem of $\Z_N$-twisted $S^1$ reduction is non-trivial. It would be interesting to go over the list of known generalized AD theories \cite{Xie:2012hs,Cecotti:2011rv,Wang:2015mra,Wang:2018gvb,Beem:2023ofp} and determine what happens to them under such a reduction. In particular, which ones obey the non-integrality and thus lose their Coulomb branch after the $\Z_N$-twisted reduction. And for those that violate the non-integrality, determine the portion of their Coulomb branch that survives, and what moduli space one gets in the 3d limit. These all are interesting problems for the future.

The resulting 3d $\cN=4$ theory can then be A-twisted to obtain a 3d TQFT. This was the idea in \cite{Dedushenko:2018bpp}, where a connection between such a TQFT and the chiral algebra of the AD theory was observed. In this paper we have shown that indeed, this 3d TQFT possesses a natural holomorphic boundary condition $H_\varepsilon$ supporting the \emph{same} chiral algebra. This, to some extent, demystifies the observations from \cite{Dedushenko:2018bpp}, leaving further studies and rigorous investigations for the future.

Let $\cT$ be the 3d TQFT we get, and $\cC_\cT$ be its category of lines. Such lines terminating on the boundary $H_\varepsilon$ create modules for the boundary VOA $V$. Thus one obtains a functor from $\cC_\cT$ to the appropriate category of $V$-modules. It is very natural to make the following
\begin{equation*}
\text{Conjecture:} \quad \cC_\cT \rightarrow \left[V-{\rm mod}\right] \text{ is an equivalence,}
\end{equation*}
for the appropriate choice of (derived) category of $V$-module. One can choose the module category to coincide with the image of this functor, however, it remains to show that non-isomorphic lines produce non-isomorphic modules. Furthermore, there might be independent mathematically natural choices of the module category, which can be compared to the image of $\cC_\cT$. Certainly if one expects the surface operators in 4d to generate all the necessary modules of the VOA, then by the earlier argument they descend to the line operators in 3d.

One particularly nice case is when $\cC_\cT$ is a (non-unitary) Modular Tensor Category (MTC) with finitely many simple objects (which is part of the standard definition of MTC \cite{book:BK}). Then the TQFT $\cT$ follows by the famous construction \cite{Reshetikhin:1991tc,book:T} and is semi-simple. There is certainly a one-to-one correspondence between simple lines and simple V-modules in this case, which may be seen as the non-unitary incarnation of RCFT/TQFT. The prototypical example of this follows from the class of AD theories:
\begin{equation}
(A_1, A_{2n}),
\end{equation}
which have non-unitary minimal models for their VOA, and a nice MTC $\cC_{2n}$ of ordinary highest-weight modules, with finitely many simple objects. In \cite{Dedushenko:2018bpp} an evidence was given that such AD theories, after the $\Z_N$-twisted reduction (here $N=2n+3$) and the 3d A-twist, give precisely the TQFT associated with the MTC $\cC_{2n}$. In the next subsection we will see why TQFTs for such AD theories are expected to be especially nice.

More generally, $\cC_\cT$ can have infinitely many simple objects, can fail to be semisimple, or perhaps something else might go wrong. A prototypical example is the class of AD theories
\begin{equation}
(A_1, D_{2n+1}),
\end{equation}
whose VOAs are admissible-level affine $\mathfrak{su}(2)$, namely, $V_{\frac{-4n}{2n+1}}(\mathfrak{su}(2))$ \cite{Beem:2017ooy}. The appropriate module category with natural modularity properties precisely for such VOAs was constructed in \cite{Creutzig:2012sd,Creutzig:2013yca}. It is bigger than one could naively expect, and perhaps the results of these references could inform the choice of the line operator category in the 3d TQFTs obtained from the $(A_1, D_{2n+1})$ theories. (See \cite{Creutzig:2021ext} for a related study in a different TQFT, and a recent work \cite{Arakawa:2023msa} on the module categories.) Also lines in the $(A_1, A_{2p-3})$ AD theories could be studied using \cite{Auger:2019gts}. These are open questions to be addressed elsewhere. 

Now let us look at an interesting class of theories that has recently attracted some attention in the literature.

\subsection{Rank-zero 3d $\cN=4$ SCFTs}
Among the examples discussed earlier, $(A_1, A_{2n})$ stand out in that such theories have no Higgs branches. Thus after the $\Z_N$-twisted circle reduction, we obtain 3d $\cN=4$ theories that have neither Higgs nor Coulomb branch. Such theories have received some amount of attention in the recent years under the name of ``rank-zero'' 3d $\cN=4$ SCFTs \cite{Gang:2018huc,Gang:2021hrd,Gang:2023rei,Ferrari:2023fez}.

Note that in 4d $\cN=2$, the rank refers to the complex dimension of the Coulomb branch on $\R^4$. Interacting 4d $\cN=2$ rank-zero SCFTs are believed to not exist \cite{Argyres:2020nrr} (a free example being a hypermultiplet). In 3d $\cN=4$, we certainly see a plethora of interacting theories without the Coulomb branch (from the $\Z_N$-twisted circle reduction), and even without the Higgs branch, which are now called rank-zero in 3d. Such 3d rank-zero SCFTs may look surprising to the 4d $\cN=2$ practitioners, but are actually abundant.

Rank-zero theories are necessarily non-Lagrangian, in the sense that any 3d $\cN=4$ Lagrangian leads to a non-zero moduli space of vacua. In references \cite{Gang:2018huc,Gang:2021hrd,Gang:2023rei}, the authors proposed an infinite collection of 3d $\cN=2$ Lagrangians that are conjectured to flow to the rank-zero 3d $\cN=4$ SCFTs. This is analogous to the AD theories in 4d, which have no 4d $\cN=2$ Lagrangians, but 4d $\cN=1$ Lagrangian descriptions were proposed \cite{Maruyoshi:2016tqk,Maruyoshi:2016aim,Agarwal:2016pjo}.

Thus it is natural to ask what rank-zero theories follow from the $\Z_N$-twisted $S^1$ reduction of the AD theories like $(A_1, A_{2n})$. (We will conjecture an answer in the following subsection.) In fact, it makes sense to be more general and simply search for those AD theories that both obey the non-integrality and have the $C_2$-cofinite VOA \cite{Xie:2019yds,Xie:2019zlb,Xie:2019vzr} (implying the trivial Higgs branch \cite{Beem:2017ooy}). In this paper, we limit ourselves to the following set of theories:
\begin{equation}
\label{flows-to-rk0}
(A_{k-1}, A_{M-1}),\quad \text{with}\quad \gcd(k,M)=1.
\end{equation}
Such theories have no Higgs branches \cite{DelZotto:2014kka}, and it was proposed in \cite{Cordova:2015nma} that they have the $(k, k+M)$ minimal models of the $W_k$ algebra as their VOAs. The latter are $C_2$-cofinite, which agrees with the absence of the Higgs branch \cite{Beem:2017ooy}. Notice that among the examples mentioned earlier, $(A_2, A_{3n})$ and $(A_2, A_{3n-2})$ are of this type, and also obey the non-integrality. At the same time, $(A_2, A_{3n-1})$ violates both $\gcd(k,M)=1$ and the non-integrality, implying that after the twisted reduction it has both the Higgs and the Coulomb branch.

In fact, one can prove that all theories of the type \eqref{flows-to-rk0} obey the non-integrality condition. This is done using the Seiberg-Witten (SW) theory techniques elaborated for the AD theories in \cite{Xie:2012hs}. The SW curve is:
\begin{equation}
x^k + z^M=0.
\end{equation}
The allowed deformations of the SW curve are represented by the monomials:
\begin{equation}
v_{m,n} x^m z^n,
\end{equation}
where the range of possible $(m,n)$ is determined by the Newton polygon, which amounts to the following conditions \cite{Xie:2012hs}:
\begin{align}
m\geq 0,\quad n\geq 0,\quad mM + nk < kM,\cr 
m=k-1 \text{ and } n=M-1 \text{ excluded}.
\end{align}
The scaling dimensions of $x$ and $z$ are $[x]=\frac{M}{M+k}$ and $[z]=\frac{k}{M+k}$, and the scaling dimension of the deformation $v_{m,n}$ is:
\begin{equation}
\label{dim_vmn}
[v_{m,n}] = \frac{kM-mM-nk}{M+k}.
\end{equation}
The Coulomb branch operators correspond to those $v_{m,n}$ that have $[v_{m,n}]>1$. Now suppose that we found a Coulomb branch operator of integral dimension:
\begin{equation}
[v_{m,n}]=p\geq 2.
\end{equation}
Then the dimension formula \eqref{dim_vmn} implies:
\begin{equation}
kM = (p+m)M + (p+n)k.
\end{equation}
Because $p+m$ and $p+n$ are strictly positive integers, and $\gcd(k,M)=1$, this equation has no solutions. Therefore, there are no integer dimensions in the spectrum of Coulomb branch operators (or rather, generators,) and the non-integrality condition holds.

The lack of Higgs branch and the non-integrality obeyed imply that the theories \eqref{flows-to-rk0} flow to the rank-zero 3d SCFTs after the $\Z_N$-twisted circle reduction. Here the lowest common denominator of the r-charges is: 
\begin{equation}
N = M+k.
\end{equation}

This suggest an interesting \emph{open problem}: Determine the rank-zero 3d SCFTs that the $(A_{k-1}, A_{M-1})$ AD theories with $\gcd(k,M)=1$ flow to after the $\Z_{N}$-twisted circle reduction, with $N=M+k$. While our cigar construction gives the twist by $(1\mod N)\in \Z_N$, more generally, we may twist by other elements $\gamma\in\Z_N$. It was observed in the examples in \cite{Dedushenko:2018bpp} that (after the topological A-twist) this produces Galua conjugates of the $\gamma=1$ 3d TQFT.

\textbf{Remark:} We expect that the rank-zero 3d SCFTs, after the A-twist, lead to semisimple non-unitary TQFTs (which can be recovered from their MTCs by the Turaev's construction). Intuitively, the presence of the (necessarily non-compact) moduli spaces of vacua would spoil this by introducing infinite-dimensional spaces of states in the TQFT \cite{Gukov:2020lqm}. Since the A-twist produces something akin to a sigma-model into the Coulomb branch, the lifting of the Coulomb branch by the $\Z_N$ monodromy partially helps. However, the presence of the Higgs branch still leads to subtleties, like in the $(A_1, D_{2n+1})$ case, where the naive fusion rules include minus signs and a more elaborate theory is required \cite{Creutzig:2013yca}. Rank-zero theories are especially nice, since their geometric target is just a finite collection of points. These points correspond to the simple objects in the MTC, as observed in \cite{Fredrickson:2017yka,Dedushenko:2018bpp}. They are also identified with the Bethe vacua in the computation of TQFT partition functions.

\subsubsection{Conjecture for the twisted $S^1$ reduction of $(A_1, A_{2n})$}
The authors of \cite{Gang:2018huc,Gang:2021hrd,Gang:2023rei} proposed an infinite set of rank-zero 3d $\cN=4$ SCFTs. Furthermore, a compelling evidence was given in \cite{Gang:2023rei} that a certain collection of such theories labeled by a positive integer $n\geq 1$, after the topological 3d A-twist, gives non-unitary semisimple 3d TQFTs, whose RCFTs are the non-unitary $(2,2n+3)$ minimal models. These are precisely the chiral algebras of the $(A_1, A_{2n})$ AD theories. The evidence was provided by computing the TQFT partition functions using the Bethe roots technique \cite{Nekrasov:2009uh,Nekrasov:2014xaa,Closset:2017zgf,Closset:2018ghr}, and extracting the modular S and T matrices. This mimics the original computation in \cite{Dedushenko:2018bpp} (see also \cite{Kozcaz:2018usv}).

The 3d rank-zero theories $\cT_n$ in question, labeled by $n\geq 1$, are described as the $\cN=2$ abelian Chern-Simons-matter theories:
\begin{equation}
\cT_n:\quad U(1)^n_{K_n} + \text{chirals } \Phi_{a=1,\dots,n}, \text{ with the superpotential } W=V_{\mathfrak{m}_1}+\dots +V_{\mathfrak{m}_{n-1}}.
\end{equation}
Here the abelian gauge group $U(1)^n$ has the matrix of bare Chern-Simons levels:
\begin{equation}
\label{CSlevels}
K_n = 2\left( \begin{matrix}
1 & 1 & 1 & \dots & 1 & 1\\
1 & 2 & 2 & \dots & 2 & 2\\
1 & 2 & 3 & \dots & 3 & 3\\
\vdots & \vdots & \vdots & \ddots & \vdots & \vdots\\
1 & 2 & 3 & \dots & n-1 & n-1\\
1 & 2 & 3 & \dots & n-1 & n\\
\end{matrix} \right) - \frac12 {\rm Diag}(1,1,1,\dots,1).
\end{equation}
The charge matrix (i.e., the charge of $\Phi_a$ under the $b$-th $U(1)$ gauge group) is $\delta_{ab}$, so the chirals contribute $\frac12{\rm Diag}(1,\dots,1)$ to the effective CS level, canceling this term in \eqref{CSlevels}. The superpotential involves the half-BPS bare monopole operators carrying the magnetic fluxes:
\begin{align}
\mathfrak{m}_1 &= (2, -1, 0, \dots, 0),\cr
\mathfrak{m}_2 &= (-1, 2, -1, 0, \dots, 0),\cr
&\dots\cr
\mathfrak{m}_{n-1} &= (0,\dots,-1,2,-1).
\end{align}
Generic bare monopoles are not gauge-invariant in the presence of Chern-Simons levels, but these fluxes are engineered precisely in such a way that they are. It is argued in \cite{Gang:2023rei}, following the $n=1$ case in \cite{Gang:2018huc}, that the theories $\cT_n$ exhibit SUSY enhancement and indeed flow to the rank-zero 3d $\cN=4$ SCFTs.

Among these rank-zero theories, there is a minimal one corresponding to $n=1$, which also must be self-mirror. The related TQFT corresponds to the Lee-Yang MTC, as already observed in \cite{Gang:2021hrd}. The minimal theory was looked at in more depth in \cite{Ferrari:2023fez}. The \emph{loc.cit.} studied the topological 3d B-twist, not the A-twist, but that is still useful for us since the theory is expected to be self-mirror. They studied the boundary VOAs for the Dirichlet and enriched Neumann boundary conditions, and the latter appears to support the sought after $(2,5)$ minimal model, which is the VOA of the $(A_1, A_{2})$ AD theory.

The Lee-Yang TQFT was also identified in \cite{Dedushenko:2018bpp} by the $\Z_N$-twisted circle reduction and topological twist of the $(A_1, A_2)$ AD theory, where in this case $N=5$. The combined evidence strongly suggests that the minimal rank-zero 3d SCFT originates from the $(A_1, A_2)$ AD theory. More generally, we propose:
\begin{equation}
\boxed{\textbf{A conjecture:} \quad (A_1, A_{2n}) \text{ after the $\Z_{2n+3}$-twisted circle reduction flows to } \cT_n}
\end{equation}
One possible test of this conjecture would be to compute the 3d limit of the 4d superconformal index, with the $\Z_{2n+3}$ holonomy included, and compare it to the $S^3_b$ (squashed sphere) partition function of the proposed theory in 3d. This test is not performed in this paper and will be done elsewhere.

Another test is the 't Hooft anomaly matching. In the subsection \ref{sec:anom}, we computed the boundary anomaly polynomial of the $H$ (and hence $H_\varepsilon$) boundary conditions by reducing the six-form anomaly polynomial on the cigar. The answer was completely general, and it passed the consistency checks for Lagrangian theories. In particular, the coefficient in front of $p_1$ in the boundary gravitational anomaly was found to be:
\begin{equation}
-\frac{n_v-n_h}{24} = c-a.
\end{equation}
We also noted that Lagrangian boundary conditions in a Lagrangian 3d theory would have the gravitational anomaly proportional to $\frac{1}{48}$, or at worst $\frac{1}{96}$ if we break SUSY. In other words, $48(c-a)$ must be an integer (or, perhaps, a half-integer). This is definitely not the case for AD theories. For example, the $(A_1, A_2)$ theory has:
\begin{equation}
c= \frac{11}{30},\quad a=\frac{43}{120}.
\end{equation}
Thus,
\begin{equation}
48(c-a)=\frac25,
\end{equation}
which is neither an integer nor a half-integer. This seems to suggest that the $H_\varepsilon$ boundary conditions cannot be constructed using the Lagrangian tools, even if $\cT_0$ is the $\Z_5$-twisted $S^1$ reduction of the $(A_1, A_2)$. On the other hand, if the enriched Neumann boundary conditions from \cite{Ferrari:2023fez} support the $(2,5)$ minimal model, it is very tempting to suggest that they do in fact engineer the $H_\varepsilon$ boundary conditions in the IR. These two propositions are in clash.\\ \textbf{Puzzle:} What is going wrong here? Do the enriched Neumann boundary conditions not flow to $H_\varepsilon$, or do we miss something in the anomaly matching argument? This is yet another question for the future work.

\section{Relation to 2d $\cN=(0,2)$}\label{sec:2d02}
We have clarified the connection between VOA in 4d $\cN=2$ SCFT and the boundary VOA of Costello-Gaiotto in 3d $\cN=4$ theories. Let us now also connect this to 2d $\cN=(0,2)$ theories and their chiral algebras. This is done by putting our 3d TQFT on the interval, with the holomorphic boundary conditions $H_\varepsilon$ on the left and the topological boundary conditions $B$ on the right, as in Figure \ref{fig:interval}:
\begin{figure}[h]
	\centering
	\includegraphics[scale=0.5]{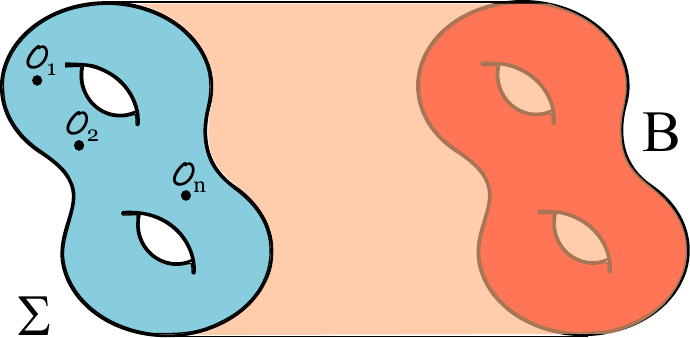}
	\caption{Introduction of the second boundary $B$, which preserves $\cN=(2,2)$ and gives the topological boundary condition in the TQFT. This configuration flows to some effective 2d QFT, which we claim preserves $\cN=(0,2)$.\label{fig:interval}}
\end{figure}\\
Using the notations from Section \ref{sec:DeformedBC}, the $\cN=(2,2)$ boundary conditions preserve supercharges $(\q_+, \bar{\q}_+, \q_-, \bar{\q}_-)$; the undeformed $\cN=(0,4)$ boundary conditions like $H$ preserve $(\q_+, \bar{\q}_+, \q_-', \bar{\q}_-')$; the deformed boundary conditions like $H_\varepsilon$ preserve $(\rQ_A^{\rm 3d}, \q_+)$, where $\rQ_A^{\rm 3d}= \bar{\q}_+ + \varepsilon \q_-$. Thus if we impose, say, $H$ on the left and some $(2,2)$ boundary condition $B$ on the right, the slab preserves $\cN=(0,2)$ supersymmetry generated by $(\q_+, \bar{\q}_+)$. If we instead impose the deformed boundary conditions $H_\varepsilon$ on the left, the slab preserves $(\q_+, \rQ_A^{\rm 3d})$. These still form the $(0,2)$ SUSY algebra, albeit embedded into the $(2,2)$ in an unusual way.

Our goal now is to determine the effective 2d $\cN=(0,2)$ description of the slab system, either with $H$ or $H_\varepsilon$ boundary conditions imposed on the left.

\subsection{Abstract generalities}\label{sec:abs-gen}
Let us explain a few general properties of the interval reduction first.

\paragraph{VOA extension.} Line operators stretched between the two boundaries, $H_\varepsilon$ and $B$, create modules for the boundary VOA $V$. In the 2d limit, the interval shrinks to a point, and lines become point-like insertions, see Figure \ref{fig:line}:
\begin{figure}[h]
	\centering
	\includegraphics[scale=0.5]{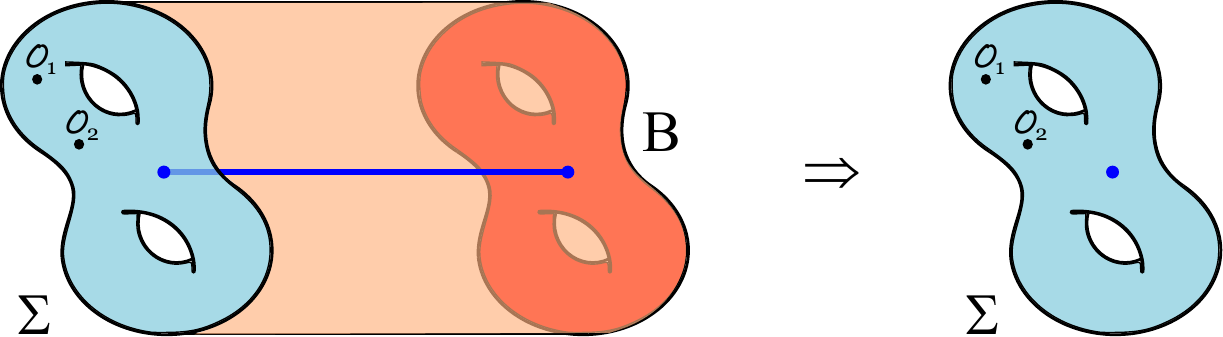}
	\caption{Blue line shrinks to a local operator in the 2d limit, which, if mutually local with the boundary VOA $V$, may extend it to a large VOA.\label{fig:line}}
\end{figure}

If these new insertions are mutually local with the vertex operators in $V$, they may extend it to a larger VOA $V_B$. In \cite{Dedushenko:2023kng,Alekseev:2022gnr} a similar interval reduction with holomorphic boundary conditions on both ends was studied, in which case the product $V_L\otimes V_R$ of the two boundary VOAs was extended by their bimodules. Here we have a simpler setup, with only one holomorphic boundary supporting $V$, leading to the extensions of $V$ itself.

The category of lines or $V$-modules determines the available extensions, but the choice of extension is determined by the boundary condition $B$, hence we use the notation $V_B$ for the extended VOA:
\begin{equation}
V\subset V_B.
\end{equation}
The bulk lines are compatible with $B$, however, some of them may be killed in the 3d limit by terminating on $B$. For example, in 3d $\cN=2$ gauge theories, the Wilson line stretched across the interval is written in terms of the complexified gauge field $\cA_y=A_y+i\sigma$, where $\sigma$ is the real adjoint scalar, and $y$ is the interval coordinate. At the $(0,2)$ Neumann boundary, $\sigma$ vanishes and $A_y$ may be set to $0$ by a gauge transformation. Thus Wilson lines with one end at such a boundary are killed in the 3d limit.

The VOA $V_B$ is the chiral algebra of the effective 2d $\cN=(0,2)$ theory obtained from the interval reduction. If such a 2d $\cN=(0,2)$ description is available, we may study $V_B$ using the two-dimensional techniques \cite{Melnikov:2019tpl,Witten:1993jg,Silverstein:1994ih,Witten:2005px,Melnikov:2009nh,Dedushenko:2015opz,Bertolini:2021hal}. Under favorable conditions, we may even hope that $V_B$ coincides with $V$. For example, when $V$ is one of the non-unitary $(2,2n+3)$ Virasoro minimal models feature earlier, it does not admit extensions by simple modules, so one expects $V_B = V$ in such cases\footnote{Note, however, that the $(2,p)$ minimal models admit logarithmic extensions \cite{Mathieu:2007ae}, hinting at the possibility of interesting phenomena.}. At the same time, $(p,q)$ minimal models with $p>2$ admit simple currents and can be extendable \cite{Mathieu:2007fx} (either as VOAs, super-VOAs, or their further generalizations).

\paragraph{Conformal blocks.} In Section \ref{sec:confblck} a state $\Psi\in \cH[\Sigma]$ had the meaning of a conformal block for the boundary VOA $V$. Here we explain that the same is true about the boundary condition $B$ in our slab setup. There are two ways to think about it.

\begin{enumerate}
	\item The boundary condition $B$ generates a boundary state $|B\rangle$. It does not belong to the physical Hilbert space since it has infinite norm. The usual remedy is to slightly evolve it in the Euclidean time, so that the state $e^{-\beta H}|B\rangle$ has finite norm and belongs to the full QFT Hilbert space.\footnote{Here it is important that the underlying QFT, before the topological twist, is unitary.} We may call $e^{-\beta H}|B\rangle$ the regularized boundary state. Now, in the topologically twisted theory, the Hamiltonian is $\rQ_A^{\rm 3d}$-exact:
	\begin{equation}
	H\propto\{\rQ_A^{\rm 3d},\dots\}.
	\end{equation}
	In particular, the kernel $\ker H$ coincides with the cohomology of $\rQ_A^{\rm 3d}$, which is the TQFT Hilbert space $\cH[\Sigma]$.	Our boundary condition preserves $\rQ_A^{\rm 3d}$, so
	\begin{equation}
	\rQ_A^{\rm 3d} |B\rangle=0,
	\end{equation}
	and the regularized boundary state $e^{-\beta H}|B\rangle$ remains $\rQ_A^{\rm 3d}$-closed, while also belonging to the physical Hilbert space. Thus it represents the physical $\rQ_A^{\rm 3d}$-cohomology class, or a vector from $\cH[\Sigma]$. We can recover this cohomology class as the projection to $\ker H$ by simply taking the limit $\beta\to +\infty$. This defines the projected state $\Pi |B\rangle$ that belongs to $\cH[\Sigma]\cong \ker H$:
	\begin{equation}
	\Pi |B\rangle := \lim_{\beta\to +\infty} e^{-\beta H}|B\rangle \in \cH[\Sigma].
	\end{equation}
	Since $\Pi|B\rangle$ manifestly belongs to the TQFT Hilbert space, it determines a conformal block for $V$. The $\rQ_A^{\rm 3d}$-exact stuff removed by $\Pi$ does not affect the BPS observables, such as the boundary correlators for $V$. Therefore, we may simply state that $|B\rangle$ itself determines a conformal block.
	\item Instead of thinking in terms of the boundary state, we can simply perform the slab reduction, resulting in a 2d $\cN=(0,2)$ theory with the chiral algebra $V_B$. As explained earlier, it extends $V$, so it contains $V$ as a sub-VOA, $V \subset V_B$. The slab decorated by the boundary insertions $\cO_1, \dots, \cO_n \in V$ reduces to the 2d theory with such insertions. The 2d correlation function can be denoted as
	\begin{equation}
	\langle \cO_1 \dots \cO_n\rangle_B.
	\end{equation}
	It determines the conformal block for $V_B$, and by restriction -- also for $V$.
\end{enumerate}

\paragraph{Boundary deformation is trivialized on the slab.} In Section \ref{sec:DeformedBC} we scrutinized the deformation $e^{\varepsilon \int (J_z^H)^{22} \dd^2 x}$ of the $(0,4)$ boundary conditions. For a theory defined on the half-space, the deformation was nontrivial, and crucial to make boundary conditions compatible with the topological 3d A-twist. Interestingly, the slab compactification trivializes this deformation. To see this, we first move the deformation term from the $(0,4)$ to the opposite $(2,2)$ boundary, which should have no effect in the 2d limit. Then we consider the effect of deformation on the $(2,2)$ boundary, and draw conclusions about the 2d limit.

Using the 3d superconformal stress-tensor multiplet, we saw in \eqref{QQJ} that $\q_+$ and $\q_-$ are preserved by this deformation, $\q_\pm (J_z^H)^{22}=0$. An important equation characterizing the deformation was \eqref{QplusJ}, but if we restrict \eqref{QplusJ} to the $(2,2)$ boundary and integrate, the right hand side vanishes. Indeed, $\int_\Sigma \dd^2z\, S_{1\perp,1\dot{2}}=0$ since it captures the non-conservation of $\q_-$, which is conserved at the $(2,2)$ boundary. The second term in \eqref{QplusJ} is still a total derivative, so it vanishes too.

Thus the deformation that used to replace $\bar{\q}_+$ by $\bar{\q}_+ + \varepsilon \q_-$ at the $(0,4)$ boundary, does nothing like that at the $(2,2)$ boundary. It simply preserves $\bar{\q}_+$, $\q_+$ and $\q_-$ there. After the slab reduction, we preserve the usual $(0,2)$ supercharges $\bar{\q}_+$ and $\q_+$ (the $(0,4)$ boundary breaks $\q_-$). Hence the deformation now has no effect on the supersymmetry: $\int(J_z^H)^{22} \dd^2 z$ must simply be some term in the action invariant under the usual $(0,2)$ SUSY.

We propose that this term in fact vanishes. This seems to be the only logical way out of the apparent contradiction: The interval-reduced theory preserves either $(\bar{\q}_+, \q_+)$ or $(\bar{\q}_+ + \varepsilon \q_-, \q_+)$, depending on whether the deformation lives on the $(2,2)$ or $(0,4)$ boundary, respectively. This argument is not very rigorous, unfortunately. However, we can explicitly check our proposal in the Lagrangian case. The boundary term is $\propto \tq^2\partial_z q^2$, while the holomorphic symplectic form for the hypermultiplets is $\delta\tq^1\wedge \delta q^1$. The $(2,2)$ boundary conditions for the hypermultiplets are determined by the complex Lagrangian submanifolds with respect to this form. The boundary term $\int_\Sigma \tq^2\partial_z q^2 \dd^2 z$ then simply vanishes.\footnote{In a 3d gauge theory, boundary conditions for the hypers are actually given by the flat Lagrangian submanifolds, such as $q^2=c$ or $\tq^2=c$, on which $\int_\Sigma \tq^2\partial_z q^2 \dd^2 z$ manifestly vanishes.}

\subsection{Lagrangian examples}
Now let us look at the Lagrangian 3d $\cN=4$ gauge theory with the gauge group $G$ and quaternionic matter representation $\mathbf{R}$. The most basic case is that of full hypermultiplets, for which one can find a gauge-invariant complex Lagrangian splitting:
\begin{equation}
\mathbf{R} = L \oplus \overline{L},
\end{equation}
where $L$ and $\overline{L}$ are dual complex representations of $G$. We think of an $\mathbf{R}$-valued hyper $H$ as a pair consisting of an $L$-valued chiral multiplet $C$ and an $\overline{L}$-valued chiral multiplet $\widetilde{C}$. The basic $(2,2)$ boundary condition is defined by giving $C$ the $(0,2)$ Neumann and $\widetilde{C}$ - the $(0,2)$ Dirichlet boundary conditions. We denote this as $D_L$. There may exist multiple inequivalent Lagrangian splittings labeling different boundary conditions for the hypers. For example, if $\mathbf{R}$ decomposes into the sum of complex irreducible representations index by $I$ as
\begin{equation}
\mathbf{R} = \bigoplus_{i\in I} \cR_i \oplus \overline{\cR}_i,
\end{equation}
then the complex Lagrangian can be chosen as
\begin{equation}
L = \bigoplus_{i\in I_L} \cR_i \oplus \bigoplus_{i\in I\setminus I_L} \overline{\cR}_i,
\end{equation}
where $I_L\subset I$ is any subset of $I$, and $L$ is labeled by a choice of such subset. In general, there may be additional labels, such as boundary vevs for the hypers (if they are compatible with gauge invariance).

For 3d $\cN=4$ vector multiplets, we use either $(2,2)$ Dirichlet or Neumann boundary conditions. Dirichlet boundary conditions break the gauge group to the global symmetry. We can also combine them by imposing the $(2,2)$ Neumann on the subgroup $H\subset G$ and the $(2,2)$ Dirichlet on the remaining $G/H$ components, like in the 4d setting of \cite{Gaiotto:2008sa}. Let us define several classes of boundary conditions in the gauge theory:
\begin{align}
\cN_L &:= \{(2,2) \text{ Neumann on the $G$ vector; $D_L$ on the hypers}\}\\
\cD_L &= \{(2,2) \text{ Dirichlet on the $G$ vector; $D_L$ on the hypers}  \}\\
\cN_{H,L}&= \{\text{Neumann on vectors in $H\subset G$, Dirichlet on $G/H$; $D_L$ on the hypers}\}
\end{align}
Notice that for $\cN_L$, $L$ must be the representation of $G$, i.e., $\mathbf{R} = L\oplus \overline{L}$ should be a $G$-invariant splitting. Thus $\cN_L$ is only defined for the full hypermultiplets. It is not defined for the half-hypers, which by definition do not possess such a splitting.

At the same time, $\cD_L$ does not require $L$ to be $G$-invariant at all, since the gauge symmetry is trivialized along the boundary. Here $L$ simply determines some complex-linear Lagrangian splitting of $\mathbf{R}$. It always exists, so $\cD_L$ is well-defined in the presence of half-hypermultiplets.

As for $\cN_{H,L}$, the splitting $L\oplus \overline{L}$ must be $H$-invariant only, it does not have to be $G$-invariant. Thus $\cN_{H,L}$ may be defined for the half-hypermultiplets valued in a pseudo-real representation $\mathbf{R}$ of $G$, as long as they become full hypermultiplets with respect to the subgroup $H$. Then the $H$-invariant splitting $L\oplus \overline{L}$ exists and $\cN_{H,L}$ is well-defined.

In addition we assume that our theory comes from the 4d $\cN=2$ superconformal gauge theory. The three-dimensional theory is fine on its own, and we could study its chiral algebra. However, our boundary conditions $H$ impose $(0,4)$ Neumann on both vector and hypermultiplets, without any boundary degrees of freedom. As we know, this only makes sense (has no gauge anomaly) when the parent 4d $\cN=2$ theory is conformal.

Now let us place this 3d theory on the interval with $H$ boundary conditions on the left and one of the $(2,2)$ boundary conditions, --- $\cD_L$, $\cN_L$, or $\cN_{H,L}$, --- on the right. Of course, we are really interested in the $H_\varepsilon$ boundary conditions, but as we explained earlier, the presence of $(2,2)$ boundary makes them equivalent to the undeformed $H$ in the IR. It is easy to determine the effective 2d theory by enumerating the multiplets of interval zero modes \cite{Dedushenko:2023kng}. The answers are:
\begin{align}
\cN_L &\Longrightarrow \boxed{\text{2d } \cN=(0,2)\ G \text{ gauge theory with chirals in }L}\\
\cD_{L} &\Longrightarrow \boxed{\text{Free }\text{2d } \cN=(0,2) \text{ chirals in }L \text{ and Fermi multiplets in }\mathfrak{g}}\\
\cN_{H,L} &\Longrightarrow \boxed{\text{2d } \cN=(0,2)\ H \text{ gauge theory with chirals in }L \text{ and Fermi in }\mathfrak{g}/\mathfrak{h}}
\end{align}

Now we may ask two things: (1) What is the chiral algebra in each of these 2d theories? (2) How is it related to the boundary VOA $V$ (which is also the VOA in 4d SCFT)?

\paragraph{Chiral algebra for $\cD_L$.} This case is the easiest to analyze since all the 2d multiplets are free. The chiral algebra in the $\bar{\q}_+$-cohomology of free chirals is the $L$-valued $\beta$-$\gamma$ system \cite{Witten:2005px}. It is also viewed as the $L\oplus\overline{L}$-valued symplectic boson, which emphasizes that it only depends on $\mathbf{R}$. We may denote it as ${\rm Sb}\otimes \mathbf{R}$. The chiral algebra in the $\bar{\q}_+$-cohomology of a free Fermi multiplet is the $bc$ ghost system \cite{Kapustin:2005pt}. Thus for $\cD_L$ we simply get:
\begin{equation}
(\mathbf{R}\text{-valued Sb})\otimes (\mathfrak{g}\text{-valued $bc$ system}).
\end{equation}

\paragraph{Chiral algebra for $\cN_L$.} Here we can only determine the perturbative chiral algebra, the non-perturbative corrections are not known. Gauging in a 2d $\cN=(0,2)$ theory, at the level of perturbative chiral algebra, corresponds to the BRST reduction \cite{Dedushenko:2017osi}. Thus for the $\cN_L$ boundary conditions imposed on the right, we get $\mathbf{R}$-valued symplectic bosons, BRST-reduced over their affine $G$ symmetry. This means computing the BRST cohomology:
\begin{equation}
H_{Q_{\rm BRST}}\big[ ({\rm Sb} \text{ in $\mathbf{R}$}) \otimes (\text{small $bc$ ghosts in $\mathfrak{g}$}) \big],
\end{equation}
where the nilpotency of $Q_{\rm BRST}$ is equivalent to cancellation of the gauge anomaly in 2d (and the conformality in 4d). Notice that the answer only depends on $\mathbf{R}$, not $L$.
The details of BRST reduction can be found in \cite{Beem:2013sza}, where it is argued that the same BRST-reduced symplectic bosons describe the chiral algebra of the parent 4d superconformal gauge theory, with the gauge group $G$ and hypermultiplets in $\mathbf{R}$. We see that it coincides with the perturbative chiral algebra in the 2d $\cN=(0,2)$ gauge theory, with the same gauge group and only half of the matter content (that is $L$, rather than the full $\mathbf{R}$). Such coincidence was first noticed in \cite{Dedushenko:2017osi}. Now we see that this is not a coincidence: The 2d gauge theory follows from the 4d one by the cigar compactification, with the $\cN_L$ boundary conditions. We will comment more on this below.

\paragraph{Chiral algebra for $\cN_{H,L}$.} This case is intermediate, sitting in between the previous two. We can also only compute the perturbative chiral algebra. It is given by the symplectic bosons in $\mathbf{R}$ and the $bc$ ghosts in $\mathfrak{g}/\mathfrak{h}$, BRST-reduced via their affine $H$ symmetry.

\paragraph{Relation to the boundary VOA $V$.} As we explained on general grounds in Section \ref{sec:abs-gen}, the 2d chiral algebra of the interval-reduced theory is the extension of the boundary chiral algebra $V$ by the line operators allowed in the compactification. Let us look through the cases considered above:
\begin{enumerate}
\item For $\cN_L$ we found that the 2d perturbative chiral algebra coincides with the boundary VOA $V$. If there are any lines in 3d that extend $V$, they must therefore correspond to the non-perturbative effects in 2d. Such effects are expected to be captured by the gauge vortices (instantons in 2d), and thus lift to some sort of vortex lines in 3d. Other lines are not allowed, for example, Wilson lines are killed by the $\cN_L$ boundary, as was mentioned before. The conclusion that for the $\cN_L$ boundary conditions,
\boxed{\text{Perturbative 2d chiral algebra = 3d boundary VOA}} is quite interesting. We expect the (non-perturbative) elliptic genus of this system to depend on $L$ and compute characters of the $V$-modules.

Note that at low energies, the 2d gauge theory roughly becomes the NLSM into the K{\"a}hler quotient $L{\mathbin{/\mkern-6mu/}}G$, which can be seen as the complex Lagrangian inside the hyper-K{\"a}hler quotient $\mathbf{R}{\mathbin{/\mkern-6mu/\mkern-6mu/\mkern-6mu/}}G$ that is the Higgs branch. Thus the perturbative VOA should be seen as some sort of chiral differential operators (CDO) \cite{1999Complex,gorbounov1999gerbes,gorbounov2001chiral} on this complex Lagrangian. In reality, these spaces are singular, and the precise statements should be about stacks and in the derived sense, as in \cite[Section 6.1]{Costello:2018swh}. Such a viewpoint may be useful for the subject of free field realizations \cite{Beem:2019tfp,Beem:2021jnm,Beem:2022vfz,Beem:2023dub} of these VOAs.
\item For $\cD_L$ and $\cN_{H,L}$ we evidently get larger (``less BRST-reduced'') algebras, with some additional $bc$ ghosts. This means that the 3d theory, even at the perturbative level, now admits lines that extend $V$. For example, Wilson lines can terminate at the $\cD_L$ boundary, and perhaps play role in the extension. It would be interesting to elucidate the details of this extension using the constructions of lines from \cite{Dimofte:2019zzj}.
\end{enumerate}

\subsection{Direct reduction from 4d}
We have considered the two-step procedure so far, first reducing from 4d $\cN=2$ to 3d $\cN=4$, and then on the interval down to 2d $\cN=(0,2)$. The topological invariance along the cigar $C$ suggests that we may shrink it to a point and reduce directly from 4d to 2d. This path may be useful for non-Lagrangian theories like the AD ones. Indeed, passing through 3d, while straightforward for the Lagrangian theories, involves a hard step in the non-Lagrangian case, which is determining the 3d limit. Furthermore, in many cases, the 4d $\cN=1$ Lagrangian for the non-Lagrangian $\cN=2$ theory is known, so one may attempt directly reducing it on the finite cigar (i.e., disk), bypassing the difficulties of the 3d limit.

Now the challenge is the choice of boundary conditions at $\Sigma\times \partial C$ in 4d. They must be the uplift of $(2,2)$ boundary conditions, thus preserving 3d $\cN=2$. This is straightforward in Lagrangian 4d $\cN=2$ theories. However, if only the 4d $\cN=1$ Lagrangian is available, we can manifestly preserve at most two supercharges at the boundary. Thus one should identify the boundary conditions that exhibit SUSY enhancement, and preserve four supercharges after the 4d $\cN=1$ bulk has flown to the $\cN=2$ point. Furthermore, the Lorentz-invariant elliptic half-BPS boundary conditions in 4d $\cN=1$ preserve 3d $\cN=1$ SUSY, which is the uplift of $\cN=(1,1)$ boundary conditions. The tip of cigar only manifestly preserves $(0,2)$ (as opposed to $(0,4)$ when the 4d has manifest $\cN=2$ SUSY). Altogether, only $(1,1)\cap (0,2)=(0,1)$ SUSY is manifest, which is not what we want. Nevertheless, could there exist non-Lorentz-invariant boundary conditions in 4d $\cN=1$, providing the 4d uplift of the $(0,2)$ boundary conditions in 3d, and preserving the same $(0,2)$ subalgebra as the tip of the cigar?

The boundary conditions in \cite{Longhi:2019hdh} seem to be of this type. However, they support infinitely many zero modes, which, if not taken care of, makes the boundary conditions unphysical. Below we explain some basics of this subtle issue, leaving the details for future work.

\subsubsection{Boundary conditions in 4d}\label{sec:4dbc}
To illustrate the problem, consider the free hypermultiplet $\cH=(Q,\tQ)$, where $Q=q+\dots$ and $\tQ=\tq + \dots$ are the $\cN=1$ chirals. The half-BPS boundary conditions (b.c.) on the individual $\cN=1$ multiplets $Q$ and $\tQ$ are labeled by real Lagrangian splittings, determined as the 3d $\cN=1$ SUSY completion of the boundary conditions like:
\begin{equation}
{\rm Re}(q)\big|=0, \quad \partial_\perp {\rm Im}(q)\big|=0,
\end{equation}
and similarly for $\tQ$. Such boundary conditions on $(Q, \tQ)$ then exhibit SUSY enhancement to 3d $\cN=2$, but this is not good enough for us. From the viewpoint of the full hypermultiplet, we could also define the 3d $\cN=2$ b.c. as the SUSY completion of:
\begin{equation}
\label{Hbc}
q\big|=0,\quad \partial_\perp \tq\big|=0,
\end{equation}
which is related to the previous b.c. by an $SU(2)_R\times SU(2)_F$ rotation of the hypermultiplet fields. Such boundary conditions, however, are not compatible with the splitting of $\cH$ into the $\cN=1$ multiplets $Q$ and $\tQ$: One can check that the SUSY completion of \eqref{Hbc} mixes fermions in the two $\cN=1$ multiplets.



This is in stark contrast with 3d. Namely, one can complete \eqref{Hbc} in 3d to the individual $\cN=(0,2)$ Dirichlet b.c. on $Q$ and $\cN=(0,2)$ Neumann b.c. on $\tQ$, and together they form the $\cN=(2,2)$ b.c. on $\cH$. These boundary conditions are elliptic in 3d. As usual, the underlying reason for such a distinction between 4d and 3d is the larger R-symmetry group in 3d. 
For us, the $(0,2)$ boundary conditions are more preferable, since they preserve the same SUSY as the tip of cigar, thus we may get the effective 2d $(0,2)$ description in the end.

What happens if we attempt to lift the $\cN=(0,2)$ Dirichlet and Neumann boundary conditions from 3d to 4d? First, the uplift cannot be Lorentz invariant any more,\footnote{Dimensional reduction of the Lorentz-invariant boundary conditions in 4d produces the non-chiral boundary conditions in 3d, such as $\cN=(1,1)$, but never $\cN=(0,2)$.} and the naive one is even non-elliptic. This is the sort of boundary conditions studied in \cite{Longhi:2019hdh}. The $(0,2)$ Dirichlet b.c. on $Q$ in 3d uplift to the Dirichlet b.c. on $Q$ in 4d that were described in \cite{Longhi:2019hdh}. We have $\R^2 \times \R_+ \times S^1$ as the 4d spacetime, where $\R_+\times S^1= C_w$ is the half-infinite cylinder with the holomorphic coordinate:
\begin{equation}
w = e^{-r + i\varphi}.
\end{equation}
The scalar $q$ obeys Dirichlet b.c. at $r=0$. Fermionic boundary conditions that follow from the 2d $(0,2)$ SUSY, though, are somewhat pathological in 4d. They admit infinitely many zero modes, labeled by the holomorphic functions $w^m$, $m\geq 0$ on the cylinder (here we assume the modes $w^m$ with $m\geq 0$ do not blow up at the infinity of $\R_+$, while those with $m<0$ do, and hence are dropped). The zero mode with $m=0$ is the bulk mode, which corresponds to the standard Kaluza-Klein reduction from 4d to 3d. This mode is the usual 2d $(0,2)$ Fermi multiplet $\Psi$ living at the standard $(0,2)$ Dirichlet b.c. on the 3d chiral multiplet (recall that such b.c. leave half of the fermions unconstrained, and they form the $(0,2)$ Fermi multiplet $\Psi$). The modes with $m>0$ are truly boundary modes. In the 3d limit, they become localized strictly at the boundary, decaying exponentially $\sim e^{- m r/R}$ as one goes into the bulk (here $R$ is the circle radius).

Analogous thing happens with the $(0,2)$ Neumann boundary conditions on the chiral $Q$ in 3d. They uplift to the so-called Robin-like boundary conditions in 4d, described in \cite{Longhi:2019hdh}. These involve (manifestly non-Lorentz-invariant) boundary conditions on the scalars:
\begin{equation}
\partial_{\bar w}q\big|=0.
\end{equation}
Such boundary conditions now carry infinitely many bosonic zero modes, corresponding to $q = w^m$, as well as their superpartners. The zero modes form the $(0,2)$ chiral multiplets $Q^{(m)}$, of which $m=0$ is the conventional bulk mode. The modes with $m>0$ again become localized at the boundary (i.e., the boundary $(0,2)$ chirals) in the strict 3d limit.

We see what is wrong with the proposed 4d uplifts of the $(0,2)$ b.c.: They are non-elliptic, and they do not reduce back to the boundary conditions in 3d that we started with. The Dirichlet boundary conditions reemerge with an infinite tower of the boundary Fermi multiplets labeled by $m=1,2,3,...$ Likewise, the Neumann boundary conditions reemerge (in the 3d limit of the Robin-like boundary conditions) accompanied by an infinite tower of the boundary chiral multiplets, also labeled by $m\geq 1$.

This also shows the path to resolve the issue of zero modes. We can simply add some boundary interactions that remove them. When we have a single chiral multiplet in the bulk, this requires adding extra boundary matter. In the case of a hyper $\cH=(Q,\tQ)$, though, the situation is especially nice. If $\tQ$ obeys the Dirichlet b.c. and $Q$ obeys the Robin b.c. in 4d, we obtain the following two towers labeled by $m\geq 1$: The boundary Fermi multiplets $\Psi^{(m)}$ originating from $\tQ$, and the boundary chirals $Q^{(m)}$ originating from $Q$. Such boundary multiplets admit symmetry-preserving masses, generated via the boundary E-terms:
\begin{equation}
\overline{D}_+ \Psi^{(m)} = E^{(m)}(Q) = Q^{(m)}.
\end{equation}
Turning on such boundary masses removes the unwanted zero modes. Indeed, if one computes the $\bT^2\times D^2$ elliptic genus of \cite{Longhi:2019hdh} for the free hyper, with the (Dirichlet,Robin) boundary conditions on $(Q,\tQ)$, all the problematic zero modes (counted by the complex structure modulus\footnote{This modulus, reminding the Omega-deformation on $D^2$, provides another way to deal with the zero modes, giving them effective masses $\propto\sigma$. This is why \cite{Longhi:2019hdh} obtain finite answers for partition functions.} denoted $\sigma$ in \cite{Longhi:2019hdh}) get canceled. One is then left with the contribution of $m=0$, which reproduces the Schur index of the hypermultiplet, as expected.

Thus there is a sense in which 4d $\cN=1$ theories admit the uplift of the $\cN=(0,2)$ boundary conditions from 3d. Such boundary conditions are necessarily non-Lorentz-invariant. Furthermore, to avoid the non-ellipticity, one is forced to include the boundary interactions removing the boundary zero modes. These details should be accounted for if we are to define the correct boundary conditions in the 4d $\cN=1$ theories flowing to the AD points.

\subsection{Sphere reduction}
Finally notice that instead of reducing the 4d theory on the cigar $C$, we could consider the sphere $S^1$ with the same $U(1)_r$ twist. Such a possibility has been considered in the literature \cite{Cecotti:2015lab}, including the most recent paper \cite{Nawata:2023aoq}. Notice that it only works nicely for theories with integral r-charges, such as Lagrangian theories. Otherwise, one would have to include defects on $S^2$, to allow for singular bundles carrying a fractional flux of $U(1)_r$. We do not investigate such a possibility and only briefly discuss Lagrangian theories.

We could view the sphere reduction in two ways: 
\begin{enumerate}
	\item We could squash the sphere into a long sausage, and then reduce it down to 3d, followed by the interval reduction to 2d. This will give us an interval with the $H$ boundary conditions at both ends. This is equivalent to having the deformed $H_\varepsilon$ boundary conditions at both ends for an even simpler reason than in our $(H,B)$ case discussed in this paper. Namely, the boundary deformation has opposite sign at the left and right boundaries (because the equation \eqref{QplusJ} concerns the outward flux of the supercharge). Therefore, the deformations simply cancel between the two boundaries. Now if we have $H_\varepsilon$ imposed at both ends of the interval, we are going to find the boundary VOA $V$ twice: one from each boundary. Then the full chiral algebra of the $S^2$-reduced theory is $V\otimes V$ extended by the bi-modules, in analogy with \cite{Alekseev:2022gnr}.
	\item We could reduce directly from 4d to 2d like in \cite{Bershadsky:1995vm}. This produces a 2d $\cN=(0,4)$ gauge theory with the hypermultiplet content $\mathbf{R}=L\oplus\overline{L}$. Its perturbative chiral algebra is the BRST reduction of: (a) one symplectic boson in $\mathbf{R}$ originating from $L$; (b) one symplectic boson in $\mathbf{R}$ originating from $\overline{L}$; (c) One adjoint-valued $bc$ system from the adjoint Fermi multiplet (a $(0,4)$ vector decomposes into the $(0,2)$ vector and the $(0,2)$ Fermi). This content will clearly produce two copies of $V$ from (a) and (b), plus a bunch of other vertex operators, which are interpreted as extending $V\otimes V$.

\end{enumerate}

\section{Outlook}\label{sec:outlook}
In this paper we tie together a bunch of topics involving Vertex (Operator) Algebras in supersymmetric QFT. We do it by utilizing the Omega-deformation formulation \cite{Oh:2019bgz,Jeong:2019pzg} of the SCFT/VOA correspondence \cite{Beem:2013sza}, and using the ideas similar to those of Nekrasov and Witten \cite{Nekrasov:2010ka}. It provides a TQFT-inspired way of thinking about the subject, connecting it to the boundary VOA in 3d $\cN=4$ \cite{Gaiotto:2016wcv,Costello:2018fnz}, and to the chiral algebras in the $\overline{Q}_+$-cohomology of 2d $\cN=(0,2)$ theories. In the first half of this paper, we work out the foundations of this approach. A lot of things are made explicit in Lagrangian theories, but the general philosophy applies to general 4d $\cN=2$ SCFTs, as we argue. We also clarify some aspects of the boundary VOA in Section \ref{sec:DeformedBC}, such as explicitly describing the boundary deformation in terms of the R-symmetry current. We then further connect our constructions to the novel subject of rank-zero 3d $\cN=4$ SCFTs \cite{Gang:2018huc,Gang:2021hrd,Gang:2023rei,Ferrari:2023fez}.

The second half lf this paper looks more like an invitation to study many open problems that the subject offers. Here we would like to list them, for reader's convenience.
\begin{itemize}
	\item Rank-zero theories from the $\Z_N$-twisted circle compactifications. Some 4d $\cN=2$ theories only have fractional r-charges in their Coulomb branch speactrum, with the lowest common denominator $N$, which results in the flavor $\Z_N$ symmetry \cite{Dedushenko:2018bpp}. After the circle KK reduction with the $\Z_N$ holonomy, the Coulomb branch of such theories disappears, which leads to interesting observations \cite{Fredrickson:2017yka,fredrickson2017s1fixed,Dedushenko:2018bpp}. Some theories, such as $(A_{k-1}, A_{M-1})$ \cite{Xie:2012hs,Cecotti:2010fi,DelZotto:2014kka} with $\gcd(k,M)=1$, while obeying this property (with $N=M+k$), also have no Higgs branch. Thus after the $\Z_N$-twisted circle reduction, they lose both branches of vacua and become the so-called rank-zero theories \cite{Gang:2018huc,Gang:2021hrd,Gang:2023rei,Ferrari:2023fez}. It is a very interesting problem to map out the landscape of such theories under the dimensional reduction; compare the set of all 3d rank-zero $\cN=4$ theories with those that originate from 4d. It is also interesting to study the $\Z_N$-twisted reduction in other theories, producing 3d SCFTs without the Coulomb branch but with the Higgs branch. Unlike the ordinary circle reduction of the Argyres-Douglas theories, such twisted reduction is completely unexplored. One can start by making conjectures and testing them, say by evaluating the appropriate limit of the 4d superconformal index, and comparing it to the $S^3_b$ partition function in 3d.
	
	\item More concretely, we conjecture that the $\Z_{2n+3}$-twisted circle reduction of the $(A_1, A_{2n})$ Argyres-Douglas theories results in the 3d rank-zero theories $\cT_n$ defined in \cite{Gang:2023rei}. It would be useful to run some tests of this conjecture. Additionally, we have run into a conundrum: The enriched Neumann boundary conditions from \cite{Ferrari:2023fez} seem to be a good candidate for (the mirror image of) $H_\varepsilon$; however, the 't Hooft anomaly-matching appears to contradict such a conclusion. Resolving this would be satisfying.
	
	\item Applying mirror symmetry in 3d, S-duality in 4d, or 2d $\cN=(2,2)$ mirror symmetry along the cigar $C$ should be interesting. On the one hand, it would relate our construction (based on the B-twist along $C$) to that of Nekrasov-Witten \cite{Nekrasov:2010ka} (based on the A-twist along $C$). On the other hand, it could possibly shed some light on the proposal of the 4d mirror symmetry/symplectic duality in \cite{Shan:2023xtw}.
	
	\item As explained in Section \ref{sec:confblck}, we have an alternative presentation of the 4d Schur index. It is captured by the 3d TQFT partition function on the solid torus with the $H_\varepsilon$ boundary conditions. One could therefore hope that such a presentation would shed some light on the Cordova-Shao formula \cite{Cordova:2015nma} expressing the Schur index in terms of the BPS spectrum on the Coulomb branch. Perhaps the generalization in \cite{Cecotti:2015lab} could also be addressed.
	
	\item The well-known problem of identifying the proper category of modules for the VOA shows up in our setup as well. With the proper choice of category of the $V$-modules (for the boundary VOA $V$), one conjecturally obtains equivalence with the category of lines in the 3d TQFT. It is interesting to explore this beyond the rational case. As mentioned in the text, it would be instructive to explore examples with the admissible level $SU(2)$. They originate from the $(A_1, D_{2n+1})$ Argyres-Douglas theories. For admissible level $SU(2)$, the category of modules obeying modularity, fusion and Verlinde formula was constructed \cite{Creutzig:2012sd,Creutzig:2013yca}. It would be fantastic to translate their construction into some kind of knowledge about the line operators in TQFT. Also the case of $(A_1, A_{2p-3})$ theories could be studied using the results of \cite{Auger:2019gts}.
	
	\item We found that an interval reduction with the boundary condition $H_\varepsilon$ on the left and $B$ on the right washes out the boundary deformation, and we get back $H$. We saw it explicitly in the Lagrangian case, but the argument given in the non-Lagrangian case perhaps could be improved.
	
	\item In the $(H,B)$ interval reduction, we find various chiral algebras, for the boundary conditions $B$ given by $\cN_L, \cD_L$, or $\cN_{H,L}$, as explained in Section \ref{sec:2d02}. Certain aspects of that calculation should be clarified, and the structure of chiral algebras as extensions of the same boundary VOA $V$ should be elucidated. In addition, the interval elliptic genera \cite{Sugiyama:2020uqh} in all these cases are expected to compute characters of the $V$-modules for the same $V$. For example, choosing the boundary conditions $\cN_L$ determined by the complex Lagrangian submanifolds $L$, we find (for different $L$'s) the same $V$ as the perturbative chiral algebra. At the non-perturbative level, though, different $L$'s should correspond to different conformal blocks of the same $V$. The (interval) elliptic genus should reveal characters of the corresponding modules.
	
	\item We obtain 2d $\cN=(0,2)$ theories by the topological reduction on the cigar $C$. While in most of this paper we take the route through 3d first, we also explain that going from 4d directly to 2d may be beneficial in some cases. This is especially interesting for the non-Lagrangian theories described via the 4d $\cN=1$ Lagrangians. In subsection \ref{sec:4dbc} we speculate on the difficulties of this approach and their possible resolution. It would be interesting to pursue it further. In particular, as a proof of concept, it would be interesting to find the 2d $(0,2)$ theory capturing the chiral algebra of the $(A_1, A_2)$ Argyres-Douglas theory.
	
	\item Such 2d $\cN=(0,2)$ models may be useful for the study of free field realizations of the VOA $V$ \cite{Beem:2019tfp,Beem:2021jnm,Beem:2022vfz,Beem:2022mde,Beem:2023dub}, since the chiral algebra of a $(0,2)$ model can sometimes be described in terms of a sheaf of VOAs on the target. It would be interesting to see if we could indeed say something new about the free field realizations using our framework.
\end{itemize}

\bibliographystyle{utphys}
\bibliography{omega-refs}
\end{document}